\documentclass[ba]{imsart}
\pubyear{2022}
\volume{TBA}
\issue{TBA}
\doi{10.1214/22-BA1313}
\arxiv{2112.03333}
\firstpage{1}
\lastpage{27}

\usepackage{amsthm}
\usepackage{amsmath}
\usepackage{amsfonts}
\usepackage{amscd}
\usepackage{mathtools}

\usepackage{natbib}
\usepackage[colorlinks,citecolor=blue,urlcolor=blue,filecolor=blue,backref=page]{hyperref}
\usepackage{graphicx}

\startlocaldefs
\usepackage{floatrow}
\newfloatcommand{capbtabbox}{table}[][\FBwidth]

\usepackage{enumerate}
\usepackage[shortlabels]{enumitem} 

\usepackage{epsfig}
\usepackage{varwidth}

\usepackage[table,dvipsnames]{xcolor}

\usepackage{booktabs}

\usepackage{bm}

\usepackage[algoruled,ruled,vlined]{algorithm2e}

\makeatletter
\renewcommand{\SetKwInOut}[2]{%
  \sbox\algocf@inoutbox{\KwSty{#2}\algocf@typo:}%
  \expandafter\ifx\csname InOutSizeDefined\endcsname\relax
    \newcommand\InOutSizeDefined{}\setlength{\inoutsize}{\wd\algocf@inoutbox}%
    \sbox\algocf@inoutbox{\parbox[t]{\inoutsize}{\KwSty{#2}\algocf@typo:\hfill}~}\setlength{\inoutindent}{\wd\algocf@inoutbox}%
  \else
    \ifdim\wd\algocf@inoutbox>\inoutsize%
    \setlength{\inoutsize}{\wd\algocf@inoutbox}%
    \sbox\algocf@inoutbox{\parbox[t]{\inoutsize}{\KwSty{#2}\algocf@typo:\hfill}~}\setlength{\inoutindent}{\wd\algocf@inoutbox}%
    \fi%
  \fi
  \algocf@newcommand{#1}[1]{%
    \ifthenelse{\boolean{algocf@inoutnumbered}}{\relax}{\everypar={\relax}}%
    {\let\\\algocf@newinout\hangindent=\inoutindent\hangafter=1\parbox[t]{\inoutsize}{\KwSty{#2}\algocf@typo:\hfill}~##1\par}%
    \algocf@linesnumbered
  }}%
\makeatother

\SetKwInOut{KwInput}{input}
\SetKwInOut{KwOutput}{output}

\usepackage{caption}
\usepackage{subcaption}
\usepackage[nameinlink]{cleveref}
\creflabelformat{equation}{#1#2#3}
\crefname{equation}{eq.}{eqs.}
\Crefname{equation}{Eq.}{Eqs.}

\newtheorem{theorem}{Theorem}

\newtheorem{proposition}[theorem]{Proposition}
\newtheorem{definition}{Definition}

\newcommand{\E}{\mathbb{E}}

\newcommand{\g}{\,\vert\,}

\newcommand{\prob}{\mathrm{P}}
\newcommand{\obs}{\mathrm{obs}}
\newcommand{\rep}{\mathrm{rep}}

\newcommand{\post}{\mathrm{post}}

\newcommand{\rmp}{\mathrm{p}}
\newcommand{\data}{\mbx_{\obs}}

\newcommand{\mbx}{\mathbf{x}}
\newcommand{\mbX}{\mathbf{X}}
\newcommand{\dd}{\, \mathrm{d}}

\newcommand{\rmin}{\mathrm{in}}
\newcommand{\rmout}{\mathrm{out}}

\newcommand{\N}{\mathbb{N}}
\newcommand{\R}{\mathbb{R}}

\newcommand{\rmA}{\mathrm{A}}
\newcommand{\rmB}{\mathrm{B}}

\DeclarePairedDelimiterX{\infdivx}[2]{(}{)}{%
  #1\;\delimsize\|\;#2%
}

\usepackage[colorinlistoftodos,
           prependcaption,
           textsize=footnotesize,
           textcolor=black,
           backgroundcolor=yellow!30,
           bordercolor=yellow!30]{todonotes}

\newcommand*\edits{\color{black}}

\usepackage{xr}
\endlocaldefs

\begin{document}


\begin{frontmatter}
\title{The Posterior Predictive Null}

\runtitle{}

\begin{aug}
\author[1]{\fnms{Gemma E.} \snm{Moran}\ead[label=e1]{gm2918@columbia.edu}},
\author[2]{\fnms{John P.} \snm{Cunningham}\ead[label=e2]{jpc2181@columbia.edu}}
\and
\author[3]{\fnms{David M.} \snm{Blei}\ead[label=e3]{david.blei@columbia.edu}}

\runauthor{}

\address[1]{Columbia Data Science Institute, Columbia University, \printead{e1}}
\address[2]{Department of Statistics, Columbia University, \printead{e2}}
\address[3]{Department of Statistics, Department of Computer Science, Columbia Data Science Institute, Columbia University, \printead{e3}}


\end{aug}

\begin{abstract}
  Bayesian model criticism is an important part of the practice of
  Bayesian statistics.  Traditionally, model criticism methods have
  been based on the predictive check, an adaptation of goodness-of-fit
  testing to Bayesian modeling and an effective method to understand
  how well a model captures the distribution of the data.  In modern
  practice, however, researchers iteratively build and develop many
  models, exploring a space of models to help solve the problem at
  hand.  While classical predictive checks can help assess each one,
  they cannot help the researcher understand how the models relate to
  each other.  This paper introduces the \emph{posterior predictive
    null check} (PPN), a method for Bayesian model criticism that
  helps characterize the relationships between models.  The idea
  behind the PPN is to check whether data from one model's predictive
  distribution can pass a predictive check designed for another model.
  This form of criticism complements the classical predictive check by
  providing a comparative tool.  A collection of PPNs, which we call a
  PPN study, can help us understand which models are equivalent and which models provide different
  perspectives on the data.  With mixture models, we demonstrate how a PPN
  study, along with traditional predictive checks, can help select the
  number of components by the principle of parsimony.  With
  probabilistic factor models, we demonstrate how a PPN study can help
  understand relationships between different classes of models, such
  as linear models and models based on neural networks. Finally, we
  analyze data from the literature on predictive checks to show how
  a PPN study can improve the practice of Bayesian model criticism. Code to replicate the results in this paper is available at \url{https://github.com/gemoran/ppn-code}.
\end{abstract}



\end{frontmatter}


\section{Introduction}
\label{sec:intro}

Bayesian model criticism is a crucial component of applied
data analysis. While designing and studying Bayesian models, the goal
of Bayesian model criticism is to understand in what ways the models fit the
data well and in what ways they fall short.

One of the main tools for Bayesian model criticism is the
\textit{predictive check}, an adaptation of goodness-of-fit
testing to Bayesian modeling \citep{B80,R84, M94, GMS96}.
Following the spirit of a goodness-of-fit test, a predictive check
first sets a \emph{reference} distribution, one that would have generated the
data if the model was true. The check then asks whether the
observed data---after applying a model-specific \emph{diagnostic} function,
such as a residual---could have plausibly arisen from that reference
distribution.  If the model passes this check, the observed data is
said to be consistent with the model. If the check fails, it suggests
that the model cannot adequately generate data
similar to the observed data; as such, the model does not provide a
useful representation of observed reality.

The most commonly used predictive check is the posterior predictive
check (PPC) \citep{G67,R84}. A PPC sets its reference distribution to be
the posterior predictive distribution of the data, and calculates the probability of the
observed data (filtered through a diagnostic function) under this distribution. A PPC captures the idea that an adequate model is one whose
posterior predictive provides a plausible distribution of the data.

Over the past decades, many researchers have innovated, refined, and
expanded Bayesian predictive checks. Some work has explored the
benefits of different reference distributions, such as the prior
predictive \citep{B80}, the posterior predictive \citep{G67,R84, M94,
  GMS96}, and combinations of the two \citep{EM06}. Other work
considers different ways to assess the adequacy of the observed
diagnostic, be it through a $p$-value, another measure of surprise
\citep{bayarri1999quantifying}, or a visual inspection \citep{GMS96,
  G04}. Still other work has addressed the problem of calibration,
trying to ensure that the Bayesian model check enjoys sensible
frequentist properties \citep{RVV00, BB00}. Finally, there is a line
of research that studies how to target the checks at specific
components of the model \citep{OH03, MS07}. This large body of work
provides a rich toolbox for criticizing a Bayesian model.

However, there is an important side of model criticism that predictive
checks do not address. In practice, rather than focus on a single model, most
Bayesian researchers posit and criticize many models
\citep{gelman2020bayesian}. Given such a
collection, predictive checks can assess each model individually, but
they cannot compare the models to each other.  Do some models capture
different aspects of the data?    Are some of
them equivalent to each other?

\begin{figure}
  \begin{center}
    \includegraphics[width=\textwidth]{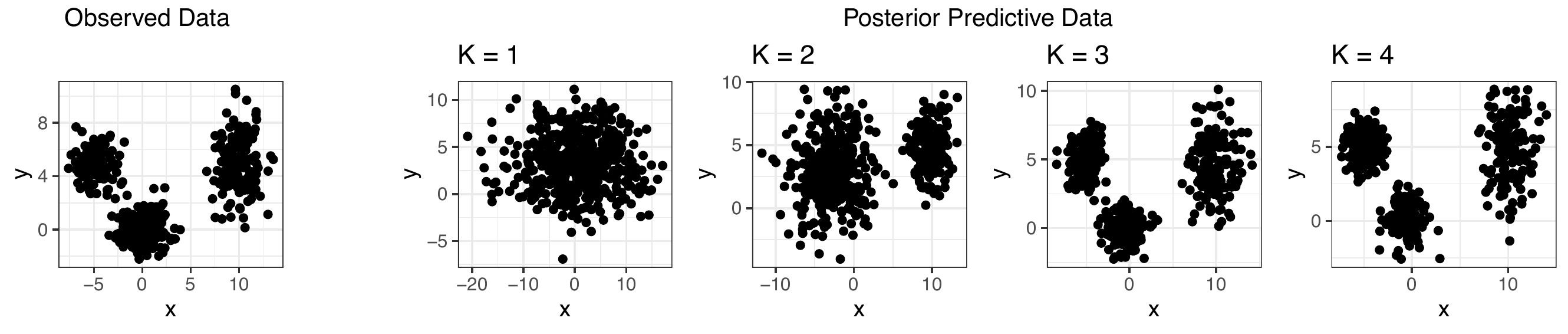}
  \end{center}
  \caption{Posterior predictive draws from Gaussian mixture models of 2D data.  On the far left is observed data $\mbx_{\obs}$ from a Gaussian mixture with $K=3$.
    Beside it are datasets drawn from the corresponding posterior
    predictive distributions of different mixtures $p_K(\mbx_{\rep}|\mbx_{\obs})$ with components
    $K \in \{1,2,3,4\}$.  Data drawn from
    $\rmp_{4}(\mbx_{\rep} \g \mbx_{\obs})$ (model-B posterior predictive) is indistinguishable from data
    drawn from $\rmp_{3}(\mbx_{\rep} \g \mbx_{\obs})$ (A-generated data).  A PPN helps
    diagnose the fact $K=4$ provides no improvement over $K=3$. (This can help a researcher choose $K=3$,  for example, on the principle of parsimony.)}
  \label{fig:gmm-ppred}
\end{figure}

To answer these questions, this paper proposes the \emph{posterior predictive null check} (PPN).
A PPN asks whether posterior predictive data from one model can pass the predictive check
of another model.  As an example, consider the simple data in \Cref{fig:gmm-ppred}
(left): two-dimensional data from a mixture of \emph{three} Gaussians. Given the
data, \Cref{fig:gmm-ppred} (right) shows the posterior predictive
distribution under four mixture models, with the number of mixture
components $K \in \{1,2,3,4\}$ (for details, see
\Cref{appendix:gmm} of the Supplementary Material). As $K$ increases, the posterior-predictive data looks
more like the true distribution of the observed data; as expected, the
posterior predictive for $K=3$ is close to the truth. But notice the
posterior predictive for $K=4$  is equally good. While we might hope
that a predictive check helps decide that $K=1$ and $K=2$ are
inadequate, how can we detect that $K=4$ offers no
improvement over $K=3$? 

The PPN helps to solve this problem, asking \emph{whether posterior
data generated from the $K=3$ posterior predictive passes the predictive check for $K=4$}. As we will discuss, answering this question is equivalent to checking whether
the predictive distribution for K = 3 is the same as the predictive distribution of $K = 4$.
If model $K=3$ passes this check, then $K=3$-generated data ``fools'' the posterior predictive check
for $K=4$; that is, the PPN suggests that $K=4$ offers no additional modeling
benefit, under the diagnostic used in the check.

\begin{figure}
  \begin{center}
    \includegraphics[width=\textwidth]{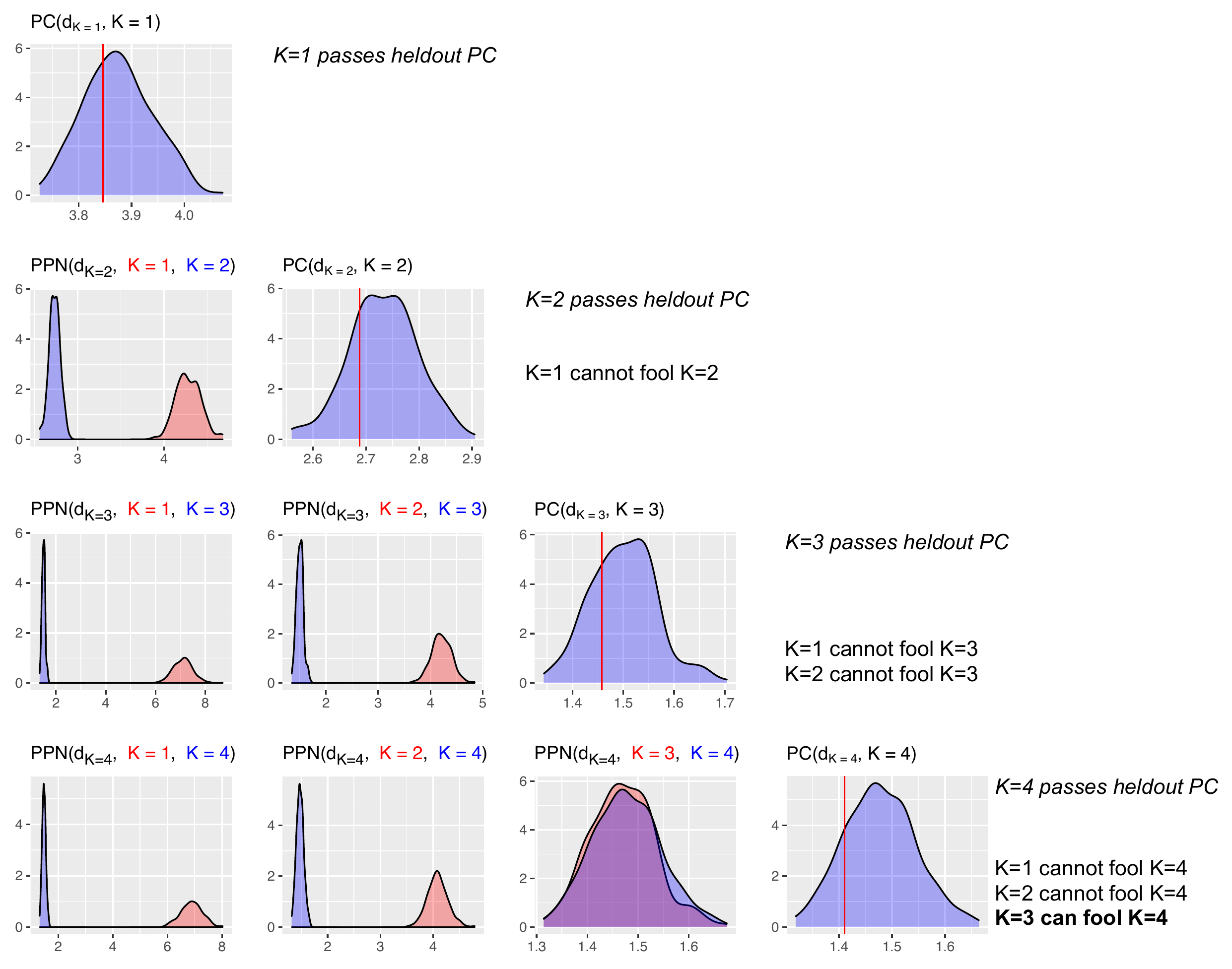}
  \end{center}
  \caption{A PPN study of mixture models which suggests $K=3$ is consistent with the data (no further mixture components are needed). The data are from \Cref{fig:gmm-ppred} (left);
    the true value of $K$ is 3. Along the diagonal are heldout
    predictive checks; every value of $K$ passes the check.  To the
    left of the diagonal are PPNs, each one checking if a simpler
    model can fool the model under study. While $K=1$ and $K=2$ pass
    their checks, the PPN shows that they cannot fool $K=3$, which
    also passes. On the other hand, $K=3$ can fool the check for
    $K=4$. }
  \label{fig:mixture-study}
\end{figure}

We note that the name PPN comes from the idea that the ``null
distribution'' of the PPC is the posterior predictive, and the
principle of the PPC is that if the data came from the null then the
model is consistent with the data. A PPN asks if an alternative
model's posterior predictive distribution might also produce the same
null distribution.

 For a set of models, a PPN study can help a researcher better understand the relationships between their models,
both how they are redundant with each other and how they differ in
their predictive distributions. As a demonstration, consider the
matrix of plots in \Cref{fig:mixture-study}.  Each row indexes a model
$K \in \{1,2,3,4\}$. Along the diagonal are classical predictive
checks---each panel illustrates the posterior predictive distribution
of the model-specific diagnostic (here, a log likelihood) and the
observed value. Notice here that all the models pass their predictive
checks; each model can expand the variance of its components to
capture the observed data. Consequently, these checks do not narrow
down the set of models under consideration.

The PPNs in the off-diagonal panels of \Cref{fig:mixture-study}
can help narrow down the set of models under consideration. Each PPN panel plots the distribution of
the diagnostic under both the model under study (the row) and a
simpler model (the column). When these distributions overlap then data
from the column's model can fool the check for the row's model.  We
see that $K=3$ cannot be fooled by $K=2$ or $K=1$; but $K=4$, though
consistent with the data, is fooled by $K=3$. When working with an
index of complexity---as we are for mixtures---the classical PPC helps
indicate if a model's complexity is sufficient to represent the
observed data, while the PPN helps to determine whether that complexity is
necessary to represent the data.  (In \Cref{sec:empirical} we will
also study collections of models that are not indexed by complexity.)

We have demonstrated how a PPN study, by appealing to the concept of
parsimony, can be used to select the number of components in a mixture
model.    We emphasize that we do not envision the PPN study as a replacement for
 model selection.  Rather, as we discussed, a PPN study can be used
as a companion to model selection methods and help to understand the relationships within a collection of
``selected'' models.  In this way, we echo the perspective of
\citet{gelman2020bayesian}, who contend that presenting multiple
models, as opposed to selecting or averaging models, provides a useful
picture of the uncertainty inherent in the process of analyzing data.  This viewpoint also connects to the ``Rashomon effect'' as coined by \citet{breiman2001statistical}: there are often many models which have equally good performance. \citet{semenova2019study} expand on this phenomena and define the ``Rashomon set'' as the set of almost equally accurate models for a given problem. A PPN study determines which models give the same posterior predictive distributions, providing a Bayesian perspective on the Rashomon effect.

In \Cref{sec:ppn} we review Bayesian predictive checks, define the
posterior predictive null check, and discuss how to use and interpret
it.  In \Cref{sec:empirical} we study and demonstrate the PPN study in
different modeling scenarios. With mixtures, we demonstrate how a PPN study can
help select the number of components.  With probabilistic factor
models, we demonstrate how a PPN study can help understand
relationships between different classes of models, such as linear
models and models based on neural networks. Finally, we analyze a
dataset from the literature on predictive checks to show how a PPN
study can enhance the practice of Bayesian model criticism.



\section{Posterior predictive null checks}
\label{sec:ppn}

\subsection{Bayesian model criticism with posterior predictive checks}

We want to analyze a dataset $\data$ with Bayesian model $\rmA$. The
model has latent variables $\theta$ and is defined by its joint,
\begin{align}
  \label{eq:model-a}
  \rmp_{\rmA}(\data, \theta_{\rmA}) =
  \rmp_{\rmA}(\data \g \theta_{\rmA})\rmp_{\rmA}(\theta_{\rmA}) .
\end{align}
Bayesian analysis proceeds by evaluating the posterior
\begin{align}
  \rmp_{\rmA}(\theta_{\rmA} \g \data) =
  \frac{\rmp_{\rmA}(\data, \theta_{\rmA})}
  {\int \rmp_{\rmA}(\mbx, \theta_{\rmA}) \dd \mbx}
\end{align}
and the corresponding posterior predictive
\begin{align}
  \label{eq:post-pred}
  \rmp_\rmA(\mbx_{\rep} \g \data) =
  \int
 \rmp_\rmA(\mbx_{\rep} \g \theta_{\rmA})\, \rmp_\rmA(\theta_{\rmA} \g \data) 
   \dd \theta_{\rmA}.
\end{align}
The posterior {\edits distribution of $\theta$} helps us investigate the latent variables; the posterior
predictive provides a distribution of new data.

Many applications of Bayesian statistics end here.  Having defined the
model, we use its posterior and posterior predictive to their intended
purposes.  But this is where the activity of Bayesian model criticism
begins.  Is the model of \Cref{eq:model-a} a good model of the data?
Does it capture the properties of the data that are important to us?
If not, in what ways does it fall short?

One of the foundational methods for Bayesian model criticism is the
\textit{posterior predictive check} (PPC), an idea that adapts
classical goodness-of-fit testing to Bayesian
statistics \citep{G67,R84}. The central premise of a PPC is that if a
model is good then its posterior predictive distribution will capture
the true distribution of the data.  If the observed data is plausible
under this predictive distribution then the model has ``passed'' the
check.  Notice that this idea takes a Bayesian approach to modeling
and a frequentist approach to checking.

There are several ingredients in a PPC.  The first is the
\textit{diagnostic statistic} $d_\rmA(\mbx)$. It is a function of
observable data that measures the incompatibility between $\mbx$
and model \textrm{A}.   As discussed in Section 4.3 of \citet{GMS96}, the choice of diagnostic should capture the aspects of the model we are interested in checking.  For example, a diagnostic which assesses the overall fitness of a model is the $\chi^2$ diagnostic:
\begin{align}
  \label{eq:residual}
  d_\rmA(\mbx) =\sum_{i=1}^n \frac{(x_i - \mathbb{E}_{\mathrm{A}}[x_i | \mbx_{\obs}])^2}{\mathrm{Var}_{\mathrm{A}}(x_i | \mbx_{\obs})},
\end{align}
where $\mathbb{E}_{\mathrm{A}}$ and $\mathrm{Var}_{\mathrm{A}}$ are expectation and variance with respect to $\mathrm{p}_{\mathrm{A}}(\mbx_\rep|\mbx_{\obs})$ (i.e. model A).  In this paper, we consider such model-dependent
diagnostic functions in order to assess whether one model can fool
another model.  In other contexts, the diagnostic might not explicitly depend on the model.

A second ingredient is the \textit{reference distribution}. When the
model is adequate, the reference distribution is the distribution of
the diagnostic $d_\rmA(\mbx)$ from which we expect the observed
diagnostic was drawn.  For their reference distribution, \citet{G67}
and \citet{R84} use the posterior predictive of the diagnostic 
$\rmp_{\rmA}(d_{\rmA}(\mbx) \g \data)$, which is derived from
\Cref{eq:post-pred}. If model $\rmA$ is a good model then the
posterior predictive of the diagnostic will capture the distribution
of the observed diagnostic.

The goal of a PPC is to evaluate whether the observed diagnostic
$d_\rmA(\data)$ could have plausibly come from the reference
distribution.  The final ingredient of a PPC is a \textit{measure of
  surprise}, a method to assess whether an observed value was drawn
from a reference.  One common approach is to use a $p$-value, a tail
probability.  A posterior predictive $p$-value is
\begin{align}
  \label{eq:post-pval}
  p_{\post} = \prob(d_\rmA(\mbx_{\rep}) \geq d_\rmA(\data) \g \data)
  && \mbx_{\rep} \sim \rmp_\rmA(\mbx_{\rep} \g \data).
\end{align}
Here a small $p$-value indicates a poor model: the observed
$d_\rmA(\data)$ is too surprising under the posterior predictive.
Note that a $p$-value is just one way to locate $d(\data)$ in its
reference distribution; graphs and other measures of surprise provide
good alternatives~\citep{GMS96, G04, BC07}.

PPCs are an intuitive method for assessing the quality of a Bayesian
model, but their statistical properties have also been criticized.
The central issue is that the PPC uses the data twice, once to
construct the reference $\rmp_\rmA(\mbx_{\rep} \g \data)$ and once to provide
the point $d_\rmA(\data)$ to locate within the reference.  The
consequence is that the PPC might be overconfident about a false
model.

\cite{BB00} and \cite{RVV00} examine this issue, both theoretically
and empirically.  They consider the sampling distribution of the
$p$-value as a function of (random) observations $\data$ from a true
likelihood $\rmp(\data \g \theta^*)$. A \textit{calibrated} $p$-value
has a uniform sampling distribution when the data truly come from this
model. Calibration is necessary to interpret $p$-values; if we do not know 
the distribution of $p$-values under the null hypothesis, we cannot make a
 decision on whether the $p$-value is ``surprising'' or not. \citet{BB00} and \citet{RVV00} show that \Cref{eq:post-pval} is not calibrated.

\cite{BB00} also propose alternative reference distributions, called partial posterior predictives, for which
the $p$-values enjoy better calibration. {\edits This paper will use an adaptation of their check, which we will refer to as the \emph{heldout 
predictive check}. The heldout predictive check divides the observed data
into two sets $\data = (\mbx_{\rmin}, \mbx_{\rmout})$, uses
$\mbx_{\rmin}$ to form the reference distribution, and locates
$d_{\rmA}(\mbx_{\rmout})$ within it. The check is
\begin{align}
  \label{eq:ppred-pval}
  p_{\mathrm{hpred}} = \prob(d_\rmA(\mbx_{\rep}) \geq d_\rmA(\mbx_{\rmout}) \g \mbx_{\rmin})
  && \mbx_{\rep} \sim \rmp_\rmA(\mbx_{\rep} \g \mbx_{\rmin}).
\end{align}
This type of check relates closely to predictive checks that rely
on cross-validation~\citep{GDC92,MS07} and held-out data~\citep{D96}.}

\subsection{Posterior predictive null checks} \label{sec:ppn-ppn}

The spirit of a predictive check is to try to falsify a model.  If we find an observed diagnostic in the tail of the reference
distribution then we ``reject the model,'' taking a $p$-value as a
measure of the risk of falsely rejecting a plausible model.  When the
observed diagnostic is not in the tail---when it has ``passed the
check''---then we have not (yet) falsified the model.  With this perspective, \cite{GS12} relate PPCs to the
Popperian view of the philosophy of science.

There is an important side of model criticism, however, that a
predictive check does not address.  Suppose model $\rmA$ is not
rejected; it passes its PPC.  This result means that $d_\rmA(\data)$
is plausible under the model-$\rmA$ predictive distribution, and we
do not reject model $\rmA$.  But does that mean we should accept
it?

To help answer this question, we propose the posterior predictive null check (PPN). Consider a different model $\rmB$ and suppose that it
provides the same posterior predictive distribution as model $\rmA$.
This means that data from model $\rmB$ will pass the predictive check
for model $\rmA$, i.e., that data from model $\rmB$ can ``fool'' the
check for model $\rmA$.  In this case, we would conclude that model $\rmB$ captures the data equally well as model $\rmA$ {\edits (with respect to the chosen diagnostic)}.  This is exactly what the PPN is designed to test.  Simply, the PPN asks whether the two models produce the same posterior predictive distribution of the model-A diagnostic.  While a predictive check assesses whether the model is adequate, a PPN helps to assess whether the model is necessary to represent the data.

Consider again \Cref{fig:gmm-ppred} (Left), which shows two-dimensional 
data from a mixture of three Gaussians. There are clearly three clusters.
\Cref{fig:gmm-ppred} (right) shows draws from the corresponding
posterior predictive for four models, $K=\{1, 2, 3, 4\}$. As expected,
a $3$-mixture provides a good posterior predictive distribution but
notice that $K=4$ does as well; it simply splits one of the clusters.
The predictive checks corroborate this visual insight---both $K=3$ and
$K=4$ pass their check, and the partial predictive $p$-values
(\Cref{eq:ppred-pval}) are 0.42 and 0.45, respectively. (In fact, each
of these models passes its check.)

A PPN can help assess $K=4$ by asking when data from the 3-mixture's posterior
predictive can fool the check for the 4-mixture.  This question amounts to asking if the
distribution of the $K=4$ diagnostic $d_{4}(\mbx_{\rep})$ is the same
whether $\mbx_{\rep}$ is drawn from $K=4$ posterior predictive, which is the
reference distribution of its predictive check, or the $K=3$ posterior predictive.
If these two posterior predictive distributions are the same then
either one will adequately locate the observed diagnostic
$d_{4}(\mbx_{\obs})$ in the $K=4$ reference distribution. Consequently, passing
the predictive check for $K=4$ does not rule out the possibility that the data came
from $K=3$ (which, in this case, it did).

\begin{definition}[Posterior predictive null check; PPN]
  Consider two models, $\rmA$ and $\rmB$, and their posterior
  predictive distributions. Each model involves its own set of latent
  variables, but defines a distribution on the same observation space $\mbx \in \mathcal{X}$,
  \begin{align*}
    \rmp_\rmA(\data, \theta_\rmA) = \rmp_{\rmA}(\data | \theta_\rmA) \rmp_{\rmA}(\theta_\rmA)
    \quad \quad \rmp_\rmA(\mbx_{\rep} \g \data) = \int \rmp_\rmA(\mbx_{\rep} \g \theta_\rmA)
    \rmp_\rmA(\theta_\rmA \g \data) \, \dd \theta_\rmA \\
    \rmp_\rmB(\data, \theta_\rmB) = \rmp_{\rmB}(\data | \theta_\rmB) \rmp_{\rmB}(\theta_\rmB)
    \quad \quad \rmp_\rmB(\mbx_{\rep} \g \data) = \int \rmp_\rmB(\mbx_{\rep} \g \theta_\rmB)
    \rmp_\rmB(\theta_\rmB \g \data) \, \dd \theta_\rmB.
\end{align*}
Consider a diagnostic function for model $\rmA$ denoted by $d_{\rmA}(\cdot)$,
such as a residual (\Cref{eq:residual}), and an observed dataset $\mbx_\obs$.  The \textit{posterior predictive null check}
$\mathrm{PPN}(d_\rmA, \rmp_\rmA, \rmp_\rmB, \mbx_\obs)$ assesses the similarity between the posterior predictive distributions of the two models under the model-$\rmA$ diagnostic.  With a divergence, $D$, the PPN is:
\begin{align}
  \label{eq:ppn-defn}
\mathrm{PPN}(d_\rmA, \rmp_\rmA, \rmp_\rmB, \mbx_\obs)=  D \infdivx{ \rmp_\rmA(d_\rmA(\mbx_{\rep}) \g \data)}{ \rmp_\rmB(d_\rmA(\mbx_{\rep}) \g \data)}.
\end{align}
\end{definition}

One example is the symmetrized Kullback-Leibler divergence:
\begin{align}
D_{\mathrm{SymKL}}\infdivx{P}{Q} = 0.5 D_{\mathrm{KL}}(P || Q) + 0.5 D_{\mathrm{KL}}(Q ||P),
\end{align}
where $D_{KL}(P||Q) = \int_{-\infty}^{\infty}   p(x) \log[p(x)/q(x)] dx$ is the Kullback-Leibler divergence between distributions $P$ and $Q$.  Another less precise example is visual inspection of the densities {\edits $\rmp_\rmA(d_{\rmA}(\mbx^\rmA_{\rep})\g \data)$ and $\rmp_\rmB(d_{\rmA}(\mbx_{\rep}^\rmB)\g\data)$ .}

Return to the Gaussian mixture model, and recall that both $K=3$ and
$K=4$ passed their respective predictive checks. We use a PPN 
to check if data from the simpler mixture ($K=3$) can fool the more
complex one ($K=4$). For the diagnostic $d_4(\cdot)$ we use the Gaussian
mixture model likelihood with $K=4$ components (for further details, 
see \Cref{appendix:gmm} of the Supplementary Material).  To implement the check,  we calculate the empirical distributions of $d_{4}(\mbx^{(3)}_{\rep})$ and $d_{4}(\mbx^{(4)}_{\rep})$, where 
$\mbx^{(3)}_{\rep}$ and $\mbx^{(4)}_{\rep}$ are draws from the posterior predictive
 of the 3- and 4-mixtures, respectively.  

This analysis is illustrated in the bottom row of \Cref{fig:mixture-study};
all the panels in the row involve the model $K=4$. In the rightmost
panel is an illustration of the partial predictive check. The distribution is the
posterior predictive of the diagnostic and the red line is the
observed diagnostic (from held-out data); the model $K=4$ passes its
predictive check. The panels to the left illustrate different PPNs, each
illustrating the predictive distribution of $\rmp(d_{4}(\mbx_{\rep})| \data, K=4)$ (blue) and $\rmp(d_{4}(\mbx_{\rep})| \data, K=k)$ for $k=1,2,3$ (red). Specifically, the leftmost panel is a PPN
that checks if data from $\rmp(\mbx_{\rep} \g \mbx_{\obs}, K=1)$ can fool the
check for $K=4$; it cannot. The next panel asks the same question for
$K=2$; again it cannot fool the check. The next panel, however,
illustrates the distribution for $K=3$; data from
$\rmp(\mbx_{\rep} \g \mbx_{\obs}, K=3)$ will pass the check for $K=4$. Thus
we cannot distinguish between the two models.  In \Cref{appendix:bayes_factors} of the Supplementary Material, we 
compare the models with Bayes factors \citep{jeffreys1998theory,kass1995bayes} and obtain a similar conclusion to the PPN study.

\subsection{The PPN study in a Bayesian workflow}

How can we incorporate the PPN study in the workflow of Bayesian data
analysis \citep{gelman2020bayesian}? First consider a set of models that are ordered by their
natural complexity. The mixture models of \Cref{fig:gmm-ppred} are a
good example. One approach to using a PPN is to iteratively increase the
complexity of the model class, use a predictive check to check each model, and
use a PPN to check whether any of the simpler models can fool the
check. Based on the principle of parsimony, one can choose the model that
passes its predictive check and for which no simpler model can fool it.

\Cref{fig:mixture-study} demonstrates this analysis for the mixture model.
First note that the predictive check does not help determine which $K$ is necessary. The diagonal plots
show how each model passes its predictive check, including the trivial model where
$K=1$; the reason is that the estimated variance in the log-likelihood is too
large to detect an anomaly between the observed and predictive
data. The off-diagonal plots demonstrate the value of the PPN study. They show that no simpler model can fool the check for
$K=3$. However, as we discussed, data from the $3$-mixture can fool
the check for the $4$-mixture. Based on this analysis, the
researcher can choose $K=3$.

Next, we consider classes
of different types of models that are not necessarily nested within
each other. We suggest using a predictive check to check each model and then use a
PPN to check every pair. This process will result in an equivalence
class of models that the data cannot distinguish. 

Consider again two models $\rmA$ and $\rmB$ and assume that they both
pass their respective predictive check.  Now consider two PPNs, one to check if
data from model $\rmB$ can fool model $\rmA$ and one to check if data
from model $\rmA$
can fool model $\rmB$.  There are three possibilities,
\begin{itemize}[leftmargin=*]
\item Suppose data from $\rmA$ fools $\rmB$ and data from $\rmB$ fools
  $\rmA$. Then these
  two models are in an equivalence class. Relative to their
  diagnostics, neither provides information that the other does
  not. We may use a qualitative criterion to select the model
  (e.g., parsimony, as we did for mixtures) or hold them both.

\item Suppose data from $\rmA$ fools $\rmB$ but data from $\rmB$ does
  not fool $\rmA$. In
  this situation, we choose model $\rmA$. It provides
  more information than model $\rmB$. (If the converse is true,
  choose model $\rmB$.)

\item Suppose data from $\rmA$ does not fool $\rmB$ and data from
  $\rmB$ does not fool
  $\rmA$. Then each model is capturing an aspect of the data
  that the other does not. Both models are valuable.
\end{itemize}

Enumerating these scenarios suggests a way to explore the differences
between classes of models, particularly those that do not necessarily
admit a natural ordering in terms of complexity.

The PPN study is related to a large literature on Bayesian predictive models for model criticism. A thorough review of such methods is provided by \citet{vehtari2012survey}.  In particular, a related method was developed in \citet{gelfand1998model}, which proposes to select a model that minimizes the expected error in predicting data from the posterior predictive distribution.  The PPN builds on such predictive methods by determining whether the posterior predictive distribution of one model can fool the predictive check for another model. In this way, the PPN provides a notion of model similarity based on predictive distributions. Further, the PPN takes into account the sampling variability in the posterior predictive distribution.



\subsection{Computing realized diagnostics}
\label{sec:realized}

The diagnostic $d_{\rmA}(\mbx)$ is a function that quantifies, in some
way, how incompatible the data $\mbx$ are to model $\rmA$. In
designing diagnostics, it is often natural and convenient to consider
function of the latent variables specified in the model. \citet{GMS96}
refer to such functions as ``realized'' because they require a
realization of the latent variables. For example, a common realized
diagnostic is the negative log likelihood of the data, \begin{align}
  d_\rmA(\mbx, \theta_\rmA) = - \log \rmp_{\rmA}(\mbx \g \theta_\rmA),
\end{align}
where $\theta_\rmA$ are the latent parameters of model $\rmA$. Large
values of this diagnostic mean the data are incompatible with the
realization of the latent variable. 

 When we use a realized diagnostic, we have to decide how to handle the
latent variable. One possibility is to remove it from the diagnostic, thereby forming a simple diagnostic from a realized one \citep{GMS96}.  Examples of such diagnostics
include the average and MAP diagnostic:
\begin{align}
  \label{gelman_marginal}
  d_{\rmA}^{\mathrm{avg}}(\mbx)
  &=
    \int
    d_{\rmA}(\mbx, \theta_{\rmA})
    \rmp(\theta_{\rmA}\g \mbx)
    \dd \theta_{\rmA} \\
  d_{\rmA}^{\rm{map}}(\mbx) &=  d_{\rmA}(\mbx, \theta_{\rmA}^*)
  \quad \quad \theta_{\rmA}^* = \arg \max_{\theta_{\rmA}} \log \rmp(\theta_{\rmA} \g \mbx).
\end{align}
Note these diagnostics can be used in the context of a PPC or a PPN.

Such diagnostics, however, 
are still computationally expensive. To evaluate each one requires a
minimization or posterior inference, and Bayesian model criticism
tends to require many evaluations of the diagnostic, one for each
replicate of the dataset.

To alleviate this burden, we propose a ``validation diagnostic.'' The
validation diagnostic marginalizes over the posterior of the latent
parameters given a fixed held-out validation dataset $\mbx_{\mathrm{val}}$, one
that is not used in the context of the model check. The validation
diagnostic is \begin{align}
  \label{eq:heldout-diagnostic}
  d_{\rmA}(\mbx ; \mbx_{\mathrm{val}})
  &=
    \int d_{\rmA}(\mbx, \theta_\rmA) \rmp_{\rmA}(\theta_\rmA \g
    \mbx_{\mathrm{val}}) \dd \theta_\rmA.
\end{align}
This diagnostic avoids the computational cost of refitting the model
to each replicated dataset. 

In practice, we split the data into
$\data = \{\mbx_{\mathrm{in}}, \mbx_{\mathrm{val}}\}$.  We use the in-sample data to draw samples from the posterior predictive distribution. The diagnostic is then defined from the
validation data.    One might ask why not use the same data in
both settings. The reason is that this would bias the diagnostic to
favor the observed data, mirroring the ``double counting'' issue of
the PPC.  Specifically, the PPN with the validation diagnostic assesses the similarity of the distributions $\rmp_\rmA(d_{\rmA}(\mbx^\rmA_{\mathrm{rep}};\mbx_{\mathrm{val}} ) \g\mbx_{\mathrm{in}})$ and $\rmp_\rmB(d_{\rmA}(\mbx^\rmB_{\mathrm{rep}};\mbx_{\mathrm{val}}  )\g \mbx_{\mathrm{in}})$.

 Note that when we use the PPN in concert with the heldout predictive check in \Cref{eq:ppred-pval}, we instead split the data into three:  $\data = \{\mbx_{\mathrm{in}}, \mbx_{\mathrm{out}}, \mbx_{\mathrm{val}}\}$, where
\begin{itemize}
	\item $\mbx_{\mathrm{in}}$ is in-sample data used to draw from the posterior predictive distribution;
	\item $\mbx_{\mathrm{out}}$ is out-of-sample data located within the reference distribution for a heldout predictive check; 
	\item $\mbx_{\mathrm{val}}$ is data used to calculate the validation diagnostic \Cref{eq:heldout-diagnostic}.
\end{itemize}
Then, the heldout predictive $p$-value with the validation diagnostic is:
\begin{align}
	\prob_{\mathrm{hpred}}(d_\rmA(\mbx_{\mathrm{rep}};\mbx_{\mathrm{val}}) \geq d_\rmA(\mbx_{\mathrm{out}}; \mbx_{\mathrm{val}}) | \mbx_{\mathrm{in}}), \quad \mbx_{\mathrm{rep}} \sim \mathrm{p}_A(\mbx_{\mathrm{rep}}| \mbx_{\mathrm{in}}).
	\label{eq:partial_pc_heldout}
\end{align}
{\edits The heldout predictive check with the validation diagnostic is in \Cref{alg:hpred}. A PPN with the validation diagnostic is in \Cref{alg:ppn_check}. Finally, a PPN study, which combines predictive checks and PPNs, is in \Cref{alg:ppn_study}.

\begin{algorithm}
\caption{The heldout predictive check} \label{alg:hpred}
\KwInput{data $\mbx_{\mathrm{obs}}=\{\mbx_{\mathrm{in}}, \mbx_{\mathrm{out}}, \mbx_{\mathrm{val}}\}$, model $\mathcal{M}_\rmA$ and diagnostic $d_\rmA(\cdot\ ;\mbx_{\mathrm{val}})$ }
\KwOutput{heldout predictive check $p$-value}
\For{$r = 1,\dots, R$}{
  draw posterior predictive data $\mbx_{\mathrm{rep},\mathrm{r}}\sim\prob(\mbx_{\mathrm{rep}}|\mbx_{\mathrm{in}}; \mathcal{M}_\rmA)$
}
\For{$b=1,\dots, B$}{
  draw samples from the posterior $\theta_{b}\sim \prob(\theta|\mbx_{\mathrm{val}};\mathcal{M}_\rmA)$
}
compute the empirical heldout predictive check $p$-value:
   \begin{align*}
   p_{hpred} &= \frac{1}{R}\sum_{r=1}^R 1\left[d_\rmA(\mbx_{\mathrm{rep}, \mathrm{r}};\mbx_{\mathrm{val}}) > d_\rmA(\mbx_{\mathrm{out}};\mbx_{\mathrm{val}})\right]\\
 \text{where } &\quad d_\rmA(\mbx;\mbx_{\mathrm{val}}) = \frac{1}{B}\sum_{b=1}^B d_\rmA(\mbx, \theta_{ b});
   \end{align*}
\KwRet{$p_{hpred}$}
\end{algorithm}

\begin{algorithm}
\caption{The posterior predictive null} \label{alg:ppn_check}
\KwInput{data $\mbx_{\mathrm{obs}}=\{\mbx_{\mathrm{in}}, \mbx_{\mathrm{out}}, \mbx_{\mathrm{val}}\}$, models $\mathcal{M}_\rmA$ and $\mathcal{M}_\rmB$ which pass their PCs, and diagnostic $d_\rmA(\cdot\ ;\mbx_{\mathrm{val}})$ }
\KwOutput{$\text{PPN}(d_\rmA, \rmp_\rmA, \rmp_\rmB, \mbx_\obs)$}
\For{$r = 1,\dots, R$}{
  draw posterior predictive data $\mbx_{\mathrm{rep},\mathrm{r}}^\rmA\sim\prob_\rmA(\mbx_{\mathrm{rep}}|\mbx_{\mathrm{in}}; \mathcal{M}_\rmA)$
}
\For{$r = 1,\dots, R$}{
  draw posterior predictive data $\mbx_{\mathrm{rep},\mathrm{r}}^\rmB\sim\prob_\rmA(\mbx_{\mathrm{rep}}|\mbx_{\mathrm{in}}; \mathcal{M}_\rmB)$
}
\For{$b=1,\dots, B$}{
  draw samples from the posterior $\theta_{b}\sim \prob(\theta|\mbx_{\mathrm{val}};\mathcal{M}_\rmA)$
}
compute the empirical PPN 
   \begin{align*}
   \text{PPN}(d_\rmA, \rmp_\rmA, \rmp_\rmB, \mbx_\obs)&=D \left(\left\{ d_\rmA(\mbx_{\rep, \mathrm{r}}^\rmA;\mbx_{\mathrm{val}})\right\}_{r=1}^R \bigg|\bigg| \left\{d_\rmA(\mbx_{\rep, \mathrm{r}}^\rmB;\mbx_{\mathrm{val}})\right\}_{r=1}^R\right)\\
   \text{where } &\quad d_\rmA(\mbx;\mbx_{\mathrm{val}}) = \frac{1}{B}\sum_{b=1}^B d_\rmA(\mbx, \theta_{ b});
   \end{align*}
\KwRet{$\text{PPN}(d_\rmA, \rmp_\rmA, \rmp_\rmB, \mbx_\obs)$}
\end{algorithm}

\begin{algorithm}
\caption{A PPN study} \label{alg:ppn_study}
\KwInput{ $\mbx_{\mathrm{obs}}=\{\mbx_{\mathrm{in}}, \mbx_{\mathrm{out}}, \mbx_{\mathrm{val}}\}$, models $\{\mathcal{M}_k\}_{k=1}^K$, diagnostics $\{d_k(\cdot\ ;\mbx_{\mathrm{val}})\}_{k=1}^K$}
\KwOutput{A collection of PPNs}
$S=\emptyset$\;
\For{$k=1,\dots, K$}{
  compute the empirical heldout predictive check for $\mathcal{M}_k$ (\Cref{alg:hpred})\;
  \If{$\mathcal{M}_k$ passes the check}{
  $S\leftarrow S\cup \{k\}$
  }
}
\For{$k \in S$}{
  \For{$j \in S \backslash \{k\}$}{
  compute $\text{PPN}(d_k, \rmp_k, \rmp_j, \mbx_\obs)$ (\Cref{alg:ppn_check})\;
  }
}
\end{algorithm}
}

\subsection{A simple regression example}
{\edits

As a simple pedagogical example of a PPN, we compare two
regression models for which the posterior predictive distributions are known exactly.  The first ``regression'' $\mathcal{M}_A$ does not include any covariates, while the second $\mathcal{M}_B$ includes $p$ covariates,
\begin{align}
y_i |\theta, \mathcal{M}_A &\sim N(\theta, 1), \quad p(\theta)\propto 1, \quad i = 1,\dots, n \label{eq:model_A_regression}\\
y_i |\theta, \beta, \mbx_i, \mathcal{M}_B&\sim N(\theta + \mbx_i^{\top}\beta, 1), \quad p(\theta, \beta)\propto 1,
\end{align}
with $\theta\in\mathbb{R}$ and $\beta, \mbx_i \in \mathbb{R}^p$.

Given observed data $(\bm{y}_{\obs}, \mbX_{\obs})$, a PPN study helps answer the questions: Do models $\rmA$ and $\rmB$ adequately capture the data? Is model $\rmA$ sufficient to model the data or is the more complex model $\rmB$ required?  In detail, the study follows these steps:
\begin{enumerate}
\item Split the data into $(\bm{y}_\obs, \mbX_\obs) = \{(\bm{y}_{\mathrm{in}}, \mbX_{\mathrm{in}}), (\bm{y}_{\mathrm{out}},\mbX_{\mathrm{out}}), (\bm{y}_{\mathrm{val}}, \mbX_{\mathrm{val}})\}$.
\item Choose a validation diagnostic,
\begin{align}
d_{\mathcal{M}}(\bm{y};\bm{y}_{\mathrm{val}}) = \sum_{i=1}^n (y_i - \mathbb{E}[y_i | \bm{y}_{\mathrm{val}}; \mathcal{M}])^2.
\end{align}
\item Calculate the posterior predictive distributions for both $\mathcal{M}_A$ and $\mathcal{M}_B$ given the in-sample data $\bm{y}_{\mathrm{in}}$: $\rmp(\bm{y}_{\mathrm{rep}}^A | \bm{y}_{\mathrm{in}}, \mbX_{\mathrm{in}}; \mathcal{M}_A)$ and $\rmp(\bm{y}_{\mathrm{rep}}^B | \bm{y}_{\mathrm{in}}, \mbX_{\mathrm{in}}; \mathcal{M}_B)$.
\item Calculate heldout predictive $p$-values for both models,
\begin{align}
\prob(d_\rmA(\bm{y}^\rmA_{\mathrm{rep}} ; \bm{y}_{\mathrm{val}}) > d_\rmA(\bm{y}_{\mathrm{out}};\bm{y}_{\mathrm{val}}) \g \bm{y}_{\mathrm{in}},  \mbX_{\mathrm{in}}; \mathcal{M}_\rmA) \\
\prob(d_\rmB(\bm{y}_{\mathrm{rep}}^\rmB ; \bm{y}_{\mathrm{val}}) > d_\rmB(\bm{y}_{\mathrm{out}};\bm{y}_{\mathrm{val}}) \g \bm{y}_{\mathrm{in}}, \mbX_{\mathrm{in}};\mathcal{M}_\rmB).
\end{align}

\item Assuming both models pass their checks, calculate the PPN, which checks if posterior predictive data from model $\rmA$ can ``fool'' posterior predictive data from model $\rmB$ (under the diagnostic $d_\rmB$),
\begin{align}
\mathrm{PPN}(d_\rmB, \rmp_\rmA, \rmp_\rmB, \bm{y}_{\mathrm{in}})=  D \infdivx{ \rmp_\rmA(d_\rmB(\bm{y}_{\rep}^\rmA ; \bm{y}_{\mathrm{val}}) \g \bm{y}_{\mathrm{in}})}{ \rmp_\rmB(d_\rmB(\bm{y}_{\rep}^\rmB;\bm{y}_{\mathrm{val}}) \g \bm{y}_{\mathrm{in}})}.\label{eq:ppn_regression}
\end{align}
\item If the PPN passes, conclude that model $\rmA$ is consistent with the data and that model $\rmB$ is not required.
\end{enumerate}

Suppose model A is true; the covariates are not involved in producing $y$.  To demonstrate the PPN study, we generated 2,000 data points from this model (\Cref{eq:model_A_regression}, $\theta=2.5$) along with ten (meaningless) covariates.  We then ran a PPN study to compare model A and model B; the results are in \Cref{fig:regression}.

We see that both models pass the heldout predictive check, the distributions of $d_\rmB(\bm{y}_{\mathrm{rep}}^\rmA;\bm{y}_{\mathrm{val}})$ and $d_\rmB(\bm{y}_{\mathrm{rep}}^\rmB;\bm{y}_{\mathrm{val}})$ are visually very similar, and their symmetric KL is 0.24.  From this study, we would correctly conclude that model A is adequate and that the more complex model B (which still passes its check) is not needed.

In this simple situation, the PPN of \Cref{eq:ppn_regression} also has good theoretical properties. Given that model $\rmA$ is true, we can prove that the distributions of $d_\rmB(\bm{y}_{\mathrm{rep}}^\rmA;\bm{y}_{\mathrm{val}})$ and $d_\rmB(\bm{y}_{\mathrm{rep}}^\rmB;\bm{y}_{\mathrm{val}})$ are asymptotically equal; the correct model $\rmA$ can ``fool'' model $\rmB$.

\begin{proposition}\label{prop:regression}

  Suppose the data $(\bm{y}_{\obs}, \mbX_{\obs})$ is drawn from model A in \Cref{eq:model_A_regression}; the covariates do not matter.  Also assume the covariates satisfy the following condition
  \begin{align}
     \label{eq:covariate_condition} \mbx_{\mathrm{in},i}^\top[\mbX_{\mathrm{obs}}^\top\mbX_{\mathrm{obs}}]^{-1}\mbx_{\mathrm{in}, i} \to p/n \quad \text{as } n\to \infty.
  \end{align}
  Then as $n\to\infty$ and $p/n\to 0$, both $d_\rmB(\bm{y}_{\rep}^\rmA;\bm{y}_{\mathrm{val}})$ and $d_\rmB(\bm{y}_\rep^\rmB;\bm{y}_{\mathrm{val}})$ converge in distribution to $2\chi_n^2$ random variables.
\end{proposition}

\begin{proof}
See \Cref{appendix:regression} of the Supplementary Material.
\end{proof}

The PPN of the proposition compares the posterior predictive distributions of model A and model B under a model-B diagnostic. It shows that these distributions are equal in the limit as $p/n\to 0$.  As for the simulation, when the data is drawn from model A, the PPN detects that model B contains no further information. Note that the condition on the covariates in \Cref{eq:covariate_condition} may hold in a number of settings. One simple example is when the covariates are distributed as $\mbx_{\obs, i}\sim N(\bm{0}_p, \bm{I}_p)$, $i=1,\dots,n$. (With conditions,
some non-diagonal covariance matrices may also satisfy \Cref{eq:covariate_condition}.)

This PPN study required the number of regression coefficients $p$ to be much smaller than the sample size $n$. However, PPN studies are also appropriate when $p\gg n$. When $p \gg n$ in model B, the distributions of $d_\rmB(\bm{y}_{\mathrm{rep}}^\rmA;\bm{y}_{\mathrm{val}})$ and $d_\rmB(\bm{y}_\rep^\rmB;\bm{y}_{\mathrm{val}})$ will be different. This difference is due to model B overfitting the data (under the improper prior). This overfitting will be detected by a heldout predictive check. If there is alternative model that does not overfit, the PPN will detect whether the additional complexity of that model is needed.

\begin{figure}
\centering
\includegraphics[width = 0.475\textwidth]{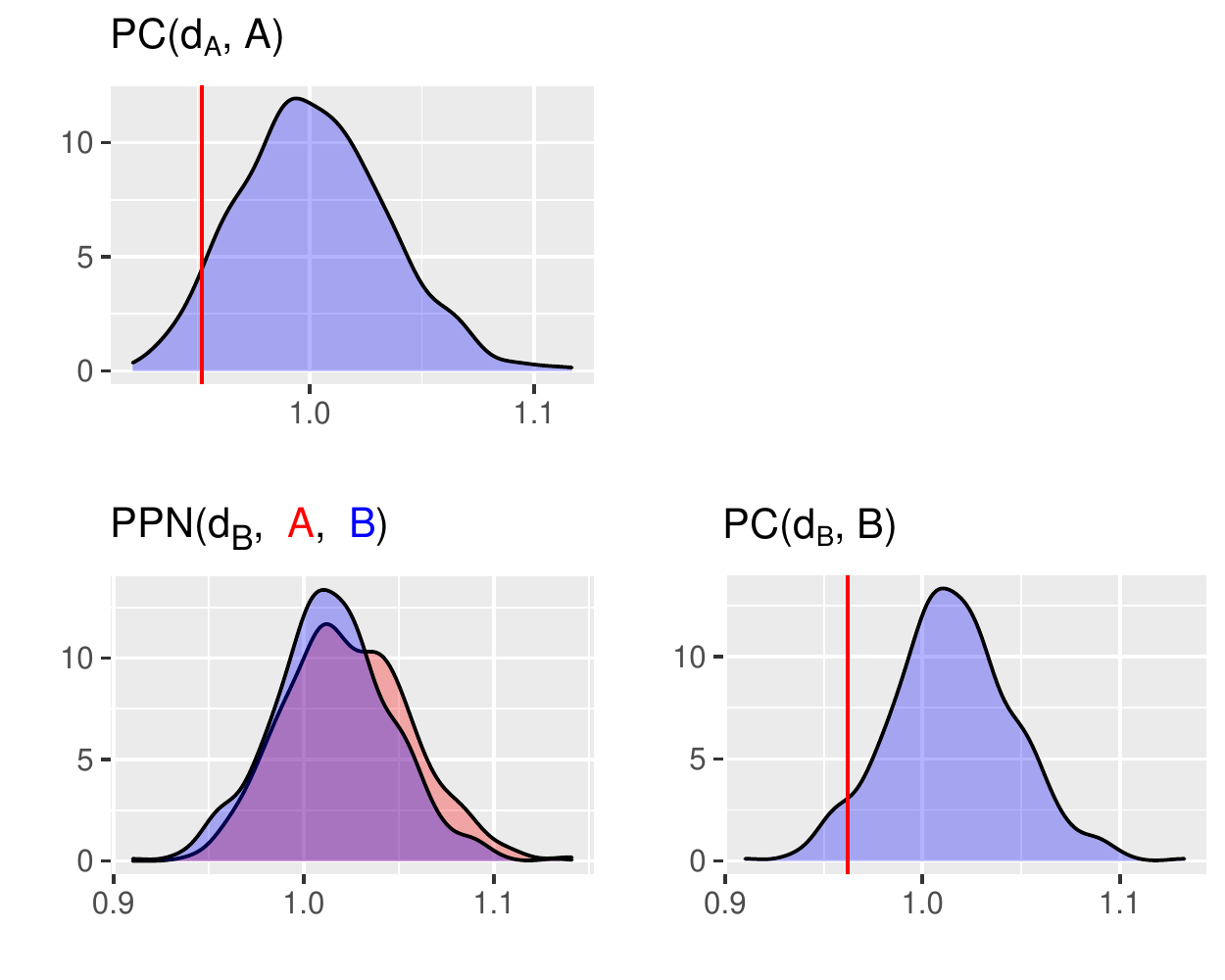}
\caption{A PPN study of regression models. This study (correctly) suggests that model $\rmA$ is consistent with the data (no covariates are needed).  On the diagonal are heldout predictive checks displaying $p_{\rmA}(d_A(\mbx_{\mathrm{rep}}^\rmA;\mbx_{\mathrm{val}})|\mbx_{\mathrm{in}})$ (blue histogram) and $d_A(\mbx_{\mathrm{out}};\mbx_{\mathrm{val}})$ (red line).  Both models pass their checks. To the left of the diagonal is a PPN which checks if model $\rmA$ can fool model $\rmB$. Specifically, the PPN compares $p_{\rmB}(d_A(\mbx_{\mathrm{rep}}^\rmB;\mbx_{\mathrm{val}})|\mbx_{\mathrm{in}})$ (red histogram) and $p_{\rmA}(d_A(\mbx_{\mathrm{rep}}^\rmA;\mbx_{\mathrm{val}})|\mbx_{\mathrm{in}})$ (blue histogram). }\label{fig:regression}
\end{figure}
}


\section{Empirical studies}
\label{sec:empirical}

We demonstrate the PPN with several empirical studies.
\begin{itemize}
	\item  \Cref{sec:gelman-example}: We consider the infant temperament data of \citet{S95}, which was also analyzed by \citet{GMS96} to illustrate PPCs with realized discrepancies.  Following the authors, we fit a multinomial mixture to the data. To choose the number of components, we use a PPN study; the result validates previous analyses of the data.
	\item  \Cref{sec:linear-example}: We consider synthetic data from a linear factor analysis model. We conduct a PPN study to choose between probabilistic PCA and two different deep generative models, fit with a variational autoencoder \citep[VAE, ][]{KW14} and skip-VAE \citep{DKRB18}, respectively. The PPN study correctly suggests that PPCA is adequate to fit the data.
	\item  \Cref{sec:nonlinear-example}: We consider synthetic data from a nonlinear factor analysis model. Here, the PPN study correctly suggests that PPCA (which assumes linearity) is \emph{not} adequate to model the data; nonlinear deep generative models provide better fits.
\end{itemize}

\subsection{Multinomial mixture model} \label{sec:gelman-example}

\citet{S95} study infant temperament data, which was also analyzed by \citet{GMS96} to illustrate PPCs with realized discrepancies.  In the study, two cohorts of infants ($n=169$, in total) were scored on the (i) degree of motor activity (scored 1-4) and (ii) crying to stimuli (scored 1-3), both at 4 months, and (iii) the degree of fear to unfamiliar stimuli at 14 months (scored 1-3).  Based on these scores, it is hypothesized that infants can be clustered into two groups: inhibited and uninhibited.  

To investigate the two-group hypothesis, we follow \citet{GMS96} and consider a multinomial mixture model.  To choose the number of mixture components, we use a PPN study.  In their analysis, \citet{GMS96} noted that ``the two-class mixture model provides an adequate fit that does not appear to improve with additional classes.'' Here the PPN study also suggests the two-class mixture model is sufficient to model the data.

For infant $i$, denote their scores in each of the three tests as $\{\mbx_{i}^{(1)}, \mbx_i^{(2)}, \mbx_i^{(3)}\}$ and their group indicator by $z_i$.  Following \citet{S95}, we assume that infants in group $k$ will have the same score probabilities, $(\theta_k^{(1)}, \theta_k^{(2)}, \theta_k^{(3)})$, across the three tests.  The multinomial mixture model with $K$ groups is:
\begin{align}
	\pi &\sim \text{Dirichlet}(\alpha_{\pi}\bm{1}_K),\\
	\theta_k^{(j)} &\sim \text{Dirichlet}(\alpha), \quad k = 1,\dots, K; \quad j = 1,2,3.\\
	z_i &\sim \text{Categorical}(\pi), \quad i = 1,\dots, n\\
	\mbx_i^{(j)} | z_i, \theta &\sim \text{Multinomial}(\theta_{z_i}^{(j)}).
\end{align}
We set $\alpha = 2$ and $\alpha_{\pi}=2$, following the recommendation of \citet{GMS96} to use a ``weak but not uniform prior distribution.'' To draw from the posterior predictive, we use Gibbs sampling.

For both partial predictive checks and PPNs, we use the heldout diagnostic \Cref{eq:heldout-diagnostic}.  The underlying diagnostic function is the $\chi^2$-discrepancy:
\begin{align}
	d_K(\mbx, \theta) &=2 \sum_{i=1}^n \sum_{k=1}^K \sum_{j=1}^3 x_{i}^{(j)} \log\left(\frac{x_{i}^{(j)}}{\E[x_{i}^{(j)} | \theta]}\right),\\
	\text{where}\quad &\E[{x}_{i}^{(j)} | \theta] = \sum_{k=1}^K  \theta^{(j)}_{k} p(z_i = k | \bm{x}_i, \theta).
\end{align}
We split the data into $\mbx = \{\mbx_{\mathrm{in}}, \mbx_{\mathrm{out}}, \mbx_{\mathrm{val}}\}$, where $\mbx_{\mathrm{in}}$ is used to draw posterior predictive data, $\mbx_{\mathrm{out}}$ is used as the out of sample data in the partial predictive check, and $\mbx_{\mathrm{val}}$ is used to define the diagnostic.  Specifically, the heldout diagnostic is:
\begin{align}
	d_K(\mbx;\mbx_{\mathrm{val}}) = \E [d_K(\mbx, \theta) | \mbx_{\mathrm{val}}],
\end{align}
which is approximated via Monte Carlo with samples from the posterior $p(\theta | \mbx_{\mathrm{val}})$.

\begin{figure}
\centering
	\begin{subfigure}{0.8\textwidth}
		\includegraphics[width =0.19\textwidth]{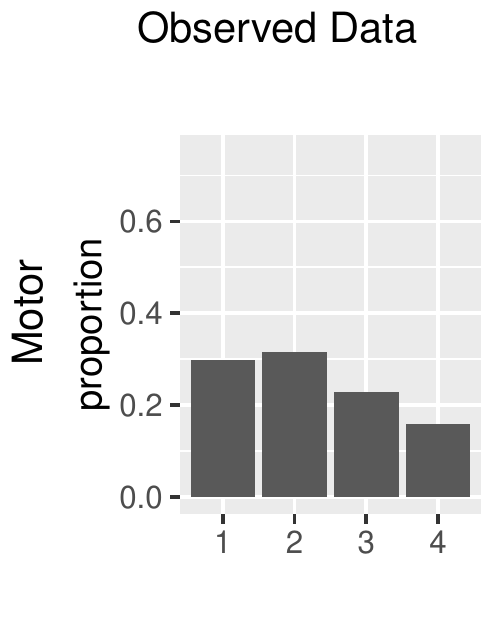}
		\includegraphics[width = 0.76\textwidth]{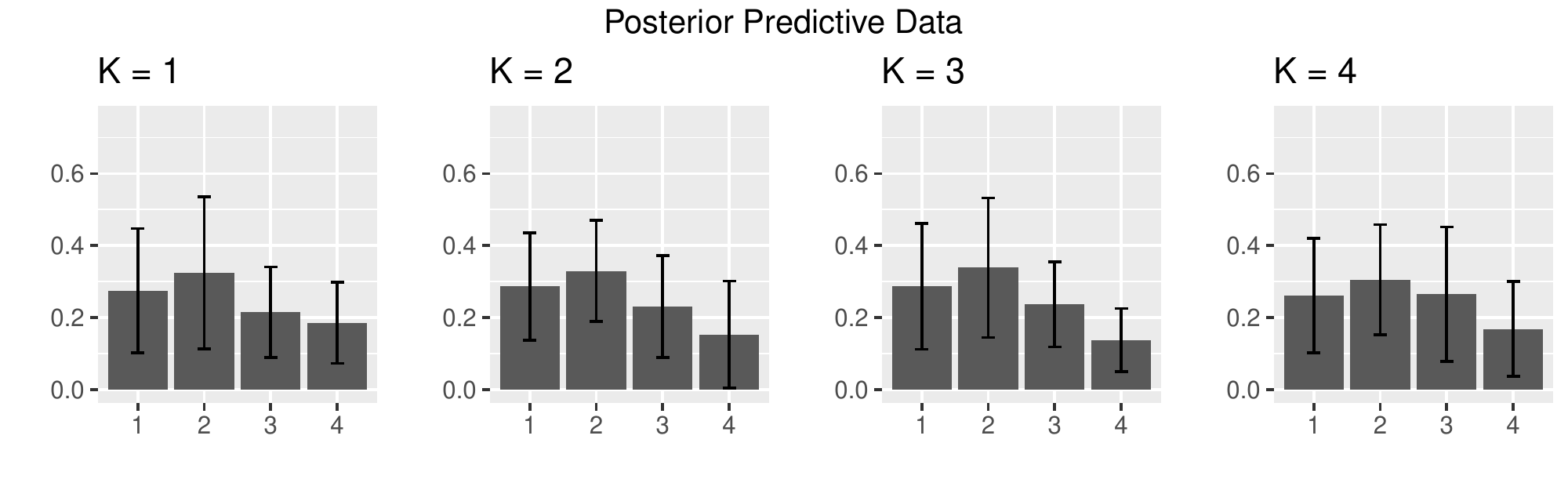}
	\end{subfigure}
	\begin{subfigure}{0.8\textwidth}
		\includegraphics[width = 0.19\textwidth]{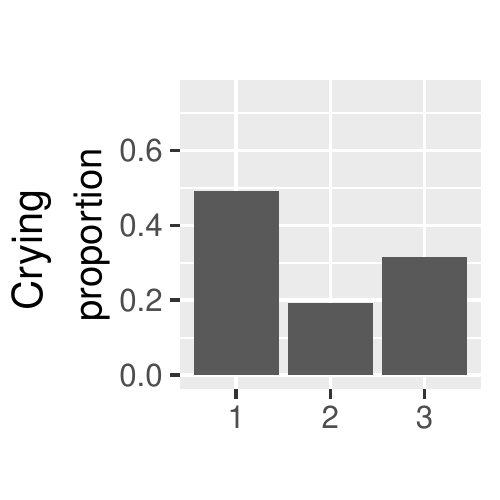}
		\includegraphics[width = 0.76\textwidth]{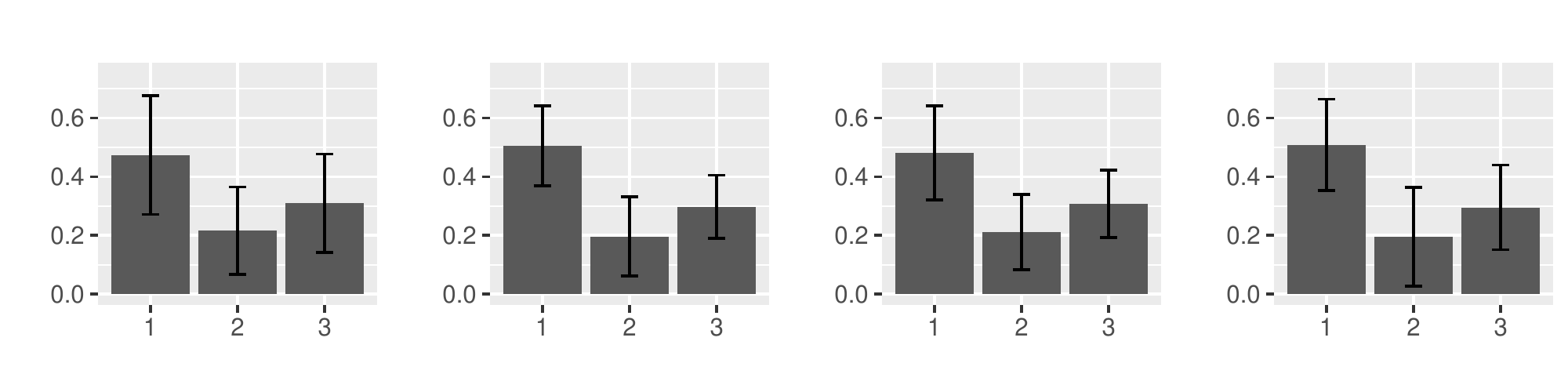}
	\end{subfigure}
	\begin{subfigure}{0.8\textwidth}
		\includegraphics[width = 0.19\textwidth]{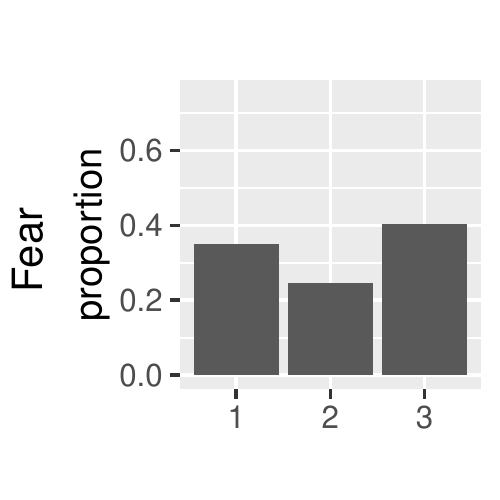}
		\includegraphics[width = 0.76\textwidth]{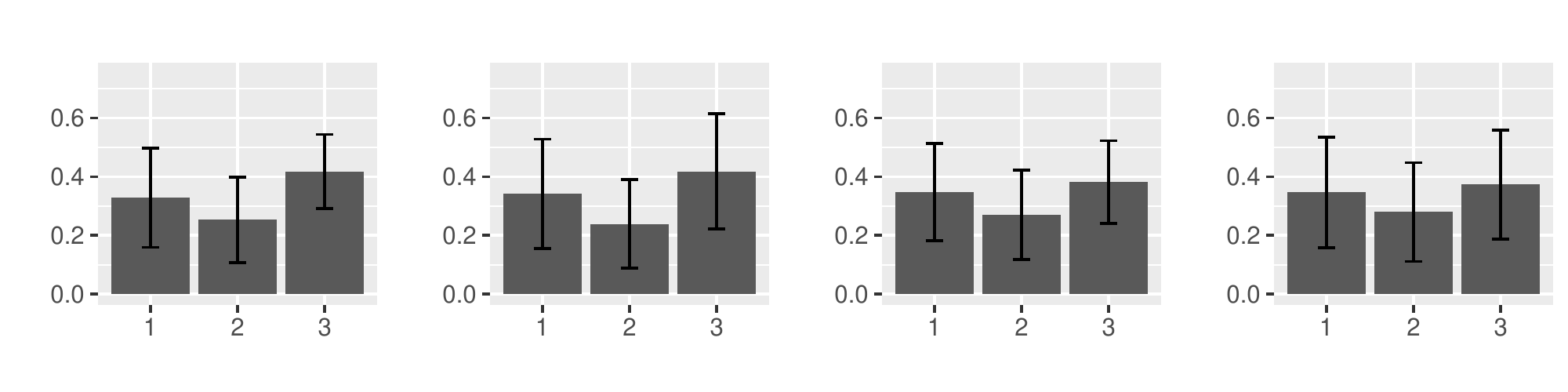}
	\end{subfigure}
\caption{Posterior predictive draws from multinomial mixture models. A PPN study will help determine which models provide no improvement over other models. On the left are the observed data proportions. Beside it are datasets draws from the posterior predictive distributions of different mixtures $p_K(\mbx_\rep|\mbx_{\mathrm{in}})$ for $K\in\{1,2,3,4\}$. Error bars are the 95\% prediction intervals. }\label{fig:gelman_pp_data}
\end{figure}

We consider mixture models with $K\in \{1, 2, 3, 4\}$ components.  All four models pass their partial predictive checks (\Cref{fig:gelman-ppn}); that is, all models generate predictive distributions which are consistent with the observed data (according to the heldout diagnostic).  As we cannot eliminate models based on goodness-of-fit, we use a PPN study to determine which models are providing essentially the same predictions. 

\noindent
\begin{minipage}{\textwidth}
\begin{minipage}[c]{0.62\textwidth}
\centering
\includegraphics[width = \textwidth]{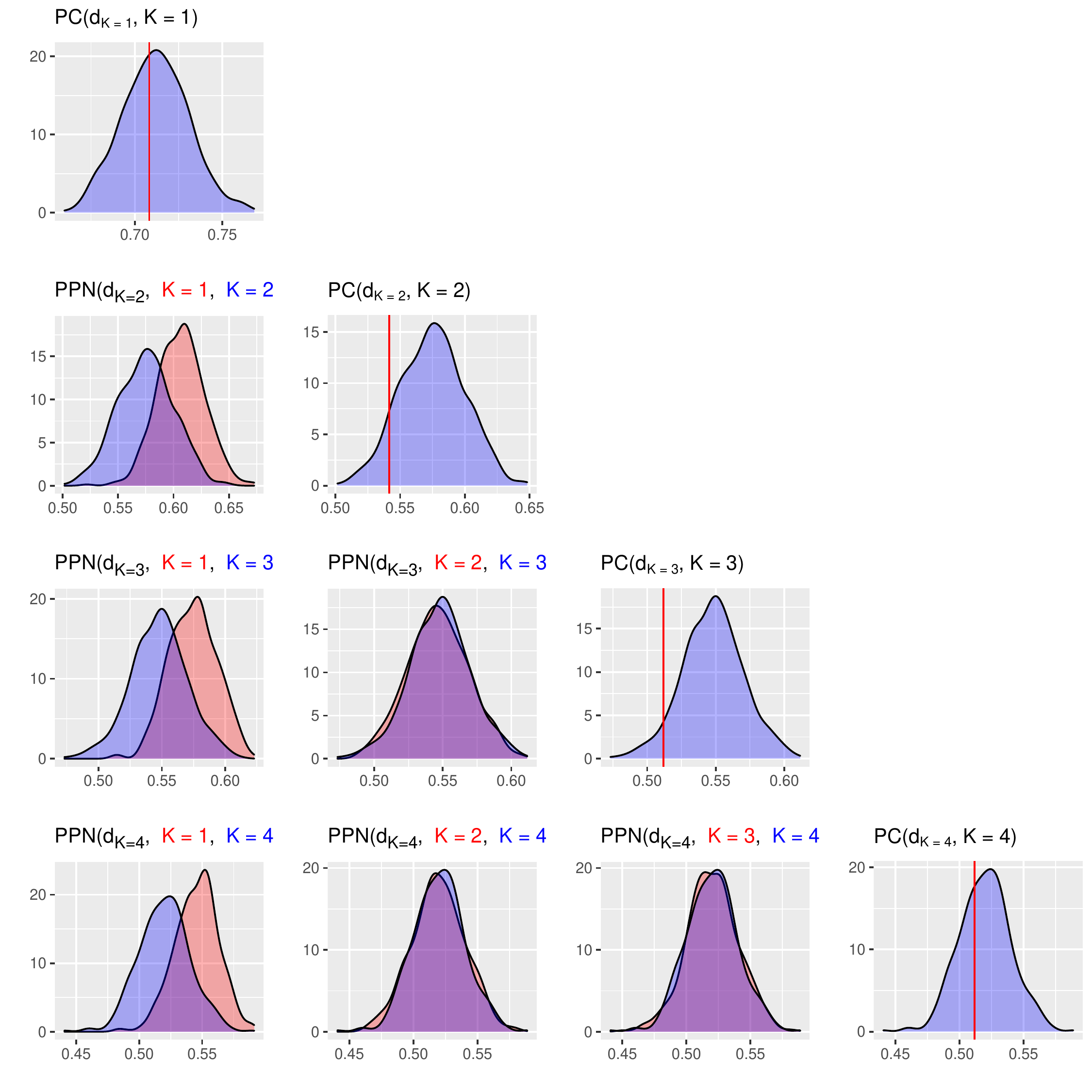}
\end{minipage}
\begin{minipage}[c]{0.33\textwidth}
\vspace{-4cm}
{\footnotesize
	\captionof{table}{Symmetrized Kullback-Leibler distance between distributions.} 
\begin{tabular}{ l r r r }
			\toprule
			& $K=1$ & $K=2$ &$K=3$  \\ 
			\midrule
$K=1$ &  &  &  \\ 
$K=  2$ &\cellcolor{gray!50} 2.18 &  &   \\ 
 $K= 3$ & \cellcolor{gray!50}2.13 &\cellcolor{gray!10} 0.16 &  \\ 
$K=  4$ & \cellcolor{gray!30}1.79 &\cellcolor{gray!10} 0.24 &\cellcolor{gray!10} 0.24   \\ 
			\bottomrule
		\end{tabular}}
\end{minipage}
\captionof{figure}{A PPN study of multinomial mixture models. This study suggests that $K=2$ is consistent with the data (no further components are needed). On the diagonal are heldout predictive checks displaying $p_{\rmA}(d_A(\mbx_{\mathrm{rep}}^\rmA;\mbx_{\mathrm{val}})|\mbx_{\mathrm{in}})$ (blue histogram) and $d_A(\mbx_{\mathrm{out}};\mbx_{\mathrm{val}})$ (red line). All models pass the checks. To the left of the diagonal are PPNs, each  one checking if a simpler model can fool the model under study. Specifically, the PPNs compare $p_{\rmB}(d_A(\mbx_{\mathrm{rep}}^\rmB;\mbx_{\mathrm{val}})|\mbx_{\mathrm{in}})$ (red histogram) and $p_{\rmA}(d_A(\mbx_{\mathrm{rep}}^\rmA;\mbx_{\mathrm{val}})|\mbx_{\mathrm{in}})$ (blue histogram).  The PPN study shows that $K=1$ passes its check but it does not fool $K=2$. $K=2$ passes its check and can fool $K=3$ and $4$.}\label{fig:gelman-ppn}
\end{minipage}

The PPN study proceeds as follows.
\begin{itemize}
\item Based on visual inspection and the symmetrized Kullback-Leibler distance, the PPN suggests that $K=1$ cannot fool $K=2$.
\item The PPN suggests that $K=2$ can fool $K=3$.
\item The PPN suggests that $K=2$ can fool $K=4$.
\end{itemize}
The PPN study suggests $K=2$ is adequate for modeling the data; increasing the number of mixture components beyond $K=2$ is not justified. This finding corroborates that of \citet{GMS96}, who made the heuristic choice of $K=2$.  The posterior predictive draws from the different models have high variability \Cref{fig:gelman_pp_data}, preventing direct visual comparison of predictive distributions at the data level. 

We additionally compute the Bayes factors to compare the models. We approximate the marginal likelihood by the harmonic mean of the likelihood values (for details, see \Cref{appendix:bayes_factors} of the Supplementary Material).  The Bayes factors provide inconclusive evidence (\Cref{table:multinomial_bf} of the Supplementary Material).

\subsection{Linear Factor Analysis} \label{sec:linear-example}

When is a nonlinear model required for factor analysis, and when is a linear model adequate?  To investigate the capacity of a PPN study to help answer this question, we consider two simulation settings.  In one, the data is generated from a linear factor model; in the other, it is generated from a nonlinear factor model.

The observed data is $\bm{x}_i \in \R^G$, $i = 1,\dots, N$. We assume that $\bm{x}_i$ has some low dimensional representation $\bm{z}_i \in \R^K$ with 
\begin{align}
	\bm{x}_i = f(\bm{z}_i) + \bm{\varepsilon}_i \label{factor_analysis_model}
\end{align}
for some function $f: \R^K \to \R^G$ and noise term $\bm{\varepsilon}_i \in \R^G$.   We consider three different modeling strategies for estimating this mapping between the latent representation and the observed data: (i) probabilistic principal components analysis \citep[PPCA, ][]{TB99}; (ii) a deep generative model, fit with a variational autoencoder \citep{KW14}; and (iii) a deep generative model with skip connections, fit with a skip-VAE \citep{DKRB18}. 

\subsubsection{Models}

Probabilistic PCA \citep{TB99} assumes that $f$ is a linear mapping from the low-dimensional latent representation to the observed data,
\begin{align*}
	\bm{x}_i = \bm{W} \bm{z}_i + \bm{\varepsilon}_i, \quad \bm{\varepsilon}_i \sim N(\bm{0}, \sigma^2\bm{I}).
\end{align*}
The latent variables are assigned a normal prior, $\bm{z}_i \sim N(0, \bm{I})$.  \citet{TB99} estimated the linear mapping $\bm{W}$ and representations $\bm{z}_i$ using the EM algorithm.

In some datasets, however, it may be that $\bm{x}_i$ lies on a much lower, nonlinear manifold. In this situation, a linear mapping would require more latent dimensions to represent the underlying structure than a nonlinear method.  To accommodate these nonlinearities, one option is to use a multi-layer feedforward neural network $\mu_{\theta}: \R^K \to \R^G$ for the mapping from the latent variables $\bm{z}_i$ to the observed data $\mbx_i$.  Such a neural network with $L\in \N$ layers has the form:
\begin{align}
\mu_{\theta}(\bm{z}) = W_{L+1}a_{b_L}\left(W_{L}a_{b_{L-1}}\cdots a_{b_1}(W_1\bm{z})\right) + b_{L+1}, \label{eq:neural_network}
\end{align}
where $p_l$ is the dimension of layer $l$, $b_l\in\R^{p_l}$ are shift vectors, $W_l \in \R^{p_l \times p_{l-1}}$ are weight matrices and $a_{b_l}$ is an activation function.  The collection of latent variables is denoted by $\theta = \{b_l, W_l\}_{l=1}^{L+1}$.

We can then model the data using this flexible neural network mapping in the following deep generative model (DGM) \citep{KW14, rezende2014stochastic}:
\begin{align*}
	\bm{z}_i &\sim N(0, \bm{I}), \\
	\bm{x}_i|\bm{z}_i &\sim N(\mu_{\theta}(\bm{z}_i), \bm{\Sigma}), \quad i = 1,\dots, N, \label{eq:DGM}
\end{align*}
where the noise variance is $\bm{\Sigma}= \text{diag}\{\sigma_j^2\}_{j=1}^G$. The latent variables $\theta$ of the neural network are generally estimated via MAP estimation. 

An alternative nonlinear mapping is a ``skip'' or residual neural network $\mu_{\theta}^{\mathrm{SKIP}}: \R^{K}\to \R^G$ \citep{DKRB18}. It includes direct connections to the latent variables $\bm{z}$ in each hidden layer of the neural network.  Specifically, the skip neural network $\mu^{\mathrm{SKIP}}_{\theta}$ has the form:
\begin{align}
\mu_{\theta}^{\mathrm{SKIP}}(\bm{z}) &= W_{L+1}^{(h)}\bm{h}_{L} + W_{L+1}^{(z)}\bm{z} \\
\text{where} \quad \bm{h}_l &= a_{b_l}(W_l^{(h)} \bm{h}_{l-1} + W_l^{(z)}\bm{z}), \quad l = 1,\dots, L, \\
\text{with}\quad \bm{h}_0 &= \bm{0}_K.
\end{align}
The skip neural network can be used in place of $\mu_{\theta}$ in the DGM in \Cref{eq:DGM}. We refer to this model as a skip-DGM.

For both the DGM and skip-DGM, posterior inference is intractable. We fit the DGM using a variational autoencoder \citep[VAE,][]{KW14, rezende2014stochastic}, which optimizes an approximation of the
regularized likelihood that uses a variational approximation of the
posterior of $p(\bm{z}_i | \bm{x})$.  The variational family is
\begin{align}
	q_{\phi}(\bm{z}_i|\bm{x}_i) \sim N(\mu_{\phi}(\bm{x}_i),  \sigma_{\phi}^2(\bm{x}_i)\odot \bm{I}), \quad i = 1,\dots, N
\end{align}
where $\mu_{\phi}: \R^G \to \R^K, \sigma^2_{\phi}: \R^G \to \R^K$ are neural networks parameterized by $\phi$ in a similar way to \Cref{eq:neural_network}.  The parameters $\theta$ and $\phi$ are estimated by optimizing the evidence lower bound (ELBO) \citep{KW14, rezende2014stochastic}. Note that the parameters $\theta$ and $\phi$ are shared across all $N$ samples and corresponding latent variables unlike mean-field variational Bayes \citep{JGJS99, BKM17}, where each sample has a unique variational parameter. This sharing of functional parameters across samples is referred to as amortized variational inference \citep{GG14}.

Similarly, we fit the skip-DGM with a skip-VAE, which includes direct connections to the observed data $\bm{x}$ in the mappings $\mu_{\phi}^{\mathrm{SKIP}}$ and $\sigma_{\phi}^{\mathrm{SKIP}}$ \citep{DKRB18}. 

For all neural networks, we use a 3-layer neural network with 20 neurons in each hidden layer.  We use a rectified linear unit (ReLU) activation in each of the hidden layers, given by $a_{b_l}(\bm{z}) = \text{max}(\bm{z} + b_l, 0)$.  To estimate the neural network parameters which maximize the ELBO, we use stochastic gradient descent with Adam \citep{KB14} with learning rate $1\times10^{-3}$. 

\subsubsection{Synthetic Data}

We first consider a linear simulation setting where we would expect PPCA to find an appropriate mapping from the latent space to the observed data, and the DGM and skip-DGM to perform similarly well.  We set the number of samples to  $N=1000$, the number of observed features to $G=10$ and the latent dimension as $K=2$. The data is generated as
\begin{align*}
	\bm{x}_i = \bm{W}\bm{z}_i + \bm{\varepsilon}_i, 
\end{align*}
where $\bm{z}_i \sim N(0, \bm{I})$, $\bm{\varepsilon}_i\sim N(0, \sigma^2\bm{I})$ with true $\sigma^2=1$.  (Note however that $\sigma^2$ is treated as unknown in the inference stage). The matrix $\bm{W}$ is the following block matrix,
\begin{align*}
	\bm{W}^{\top} = 
	\begin{pmatrix}
		5& 5 & 5 & 5 & 5 &0& 0 & 0 &0 &0 \\
		0 & 0 & 0 & 0 & 0 & 5& 5 &5&5 &5\\
	\end{pmatrix}.
\end{align*}
That is, the first five values of $\bm{x}_i$ are linearly related to the first factor, and the last five values of $\bm{x}_i$ are linearly related to the second factor. We generate three datasets: $\{\mbx_{\mathrm{in}}, \mbx_{\mathrm{out}}, \mbx_{\mathrm{val}}\}$.

\subsubsection{Model Checking}

For both the PPN study, we use a heldout diagnostic
\begin{align}
	D_{\rmA}(\mbx;\mbx_{\mathrm{val}}) = d_{\rmA}(\mbx, \widehat{\bm{\theta}}_{\mathrm{A}}), \quad \widehat{\bm{\theta}}_{\mathrm{A}}=\max_{\theta} \log p_{\rmA}(\theta | \mbx_{\mathrm{val}}).
\end{align}
That is, $\widehat{\bm{\theta}}_{\mathrm{A}}$ is the maximum \emph{a posteriori} estimate for model $\rmA$ given the validation data $\mbx_{\mathrm{val}}$. The underlying realized diagnostic, $d_{\rmA}(\mbx, \widehat{\bm{\theta}}_{\rmA})$ is the reconstruction loss,
\begin{align}
	d_{\rmA}(\mbx, \widehat{\bm{\theta}}_{\rmA}) = \sum_{i=1}^n \lVert \mbx_i - \mathbb{E}[\mbx_i| \widehat{\bm{\theta}}_{\rmA}]\rVert^2.
\end{align}

\noindent
\begin{minipage}{\textwidth}
\begin{minipage}[c]{0.49\textwidth}
\centering
\includegraphics[width =\textwidth]{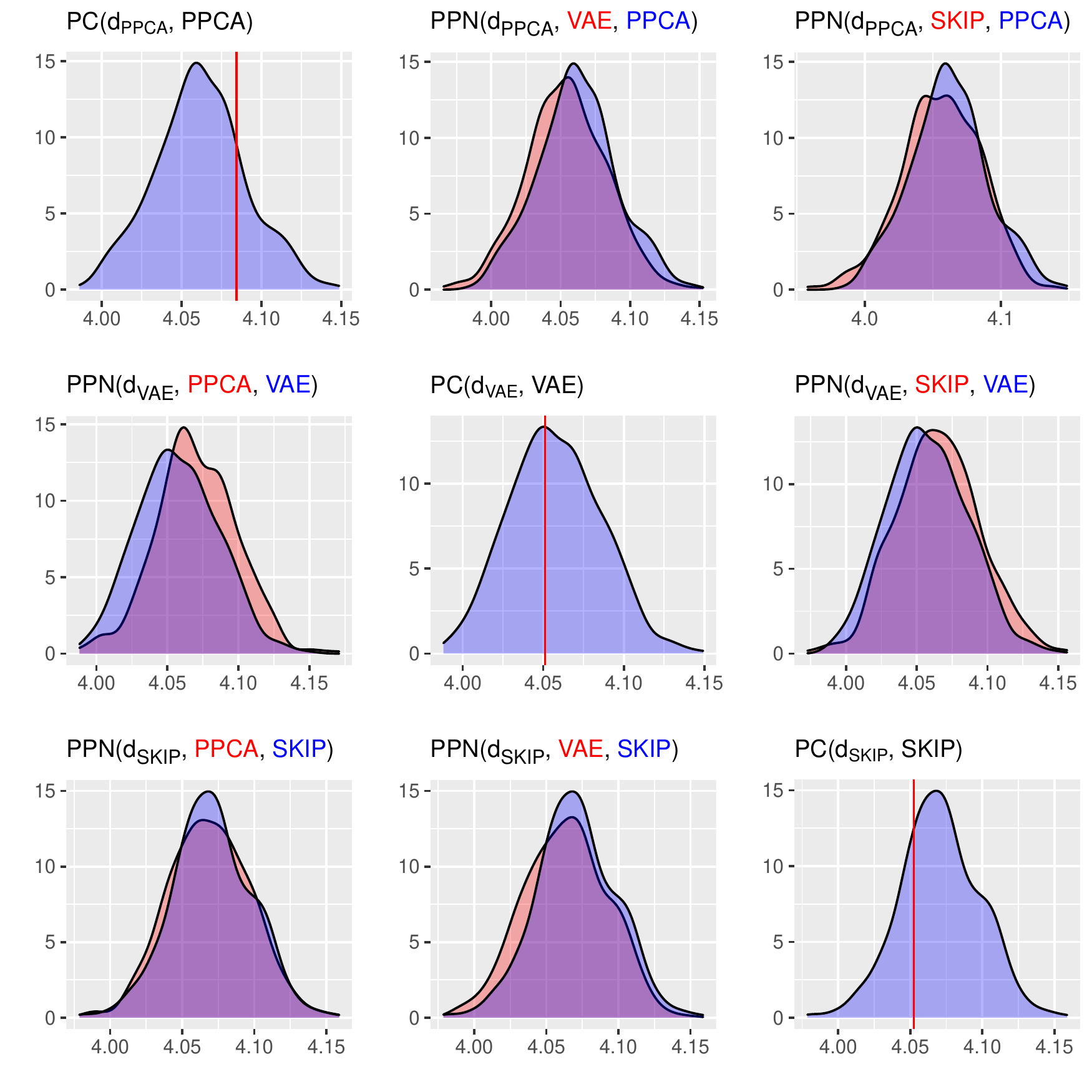}
\end{minipage}
\hfill
\begin{minipage}[c]{0.45\textwidth}
{\footnotesize
	\captionof{table}{Symmetrized Kullback-Leibler distance between distributions.}\label{table:linear} 
		\begin{tabular}{ l r r r }
		\toprule
& PPCA-2 & VAE & SKIP \\ 
\midrule
PPCA-2 & & \cellcolor{gray!10}0.22 &\cellcolor{gray!10} 0.27 \\
VAE & \cellcolor{gray!10}0.41 &  &\cellcolor{gray!10}0.25 \\
SKIP &\cellcolor{gray!10} 0.22 &\cellcolor{gray!10} 0.19 & \\
\bottomrule
		\end{tabular}}
\end{minipage}
\captionof{figure}{A PPN study of factor analysis models. This study suggests that a linear model is consistent with the data (nonlinear models are not needed). On the diagonal are heldout predictive checks displaying $p_{\rmA}(d_A(\mbx_{\mathrm{rep}}^\rmA;\mbx_{\mathrm{val}})|\mbx_{\mathrm{in}})$ (blue histogram) and $d_A(\mbx_{\mathrm{out}};\mbx_{\mathrm{val}})$ (red line). All models pass the checks. Beside the diagonal are PPNs, each one checking if a different model can fool the model under study. Specifically, the PPNs compare $p_{\rmB}(d_A(\mbx_{\mathrm{rep}}^\rmB;\mbx_{\mathrm{val}})|\mbx_{\mathrm{in}})$ (red histogram) and $p_{\rmA}(d_A(\mbx_{\mathrm{rep}}^\rmA;\mbx_{\mathrm{val}})|\mbx_{\mathrm{in}})$ (blue histogram).  The PPN study shows all models (PPCA, VAE, SKIP-VAE) fool all other models.}\label{fig:linear-ppn}
\end{minipage}

After fitting the models, each of PPCA, DGM and skip-DGM pass their partial predictive checks (\Cref{fig:linear-ppn}). It is unclear which model to choose.  The PPN study can help. Consider \Cref{fig:linear-ppn}: from both visual inspection and the symmetrized KL divergences (\Cref{table:linear}), the PPN suggests that each model fools every other model. The PPN study concludes that both PPCA and the deep generative models fit the data adequately and can be considered equivalent based on their posterior predictive distributions. If we prefer an interpretable linear model, then we should choose PPCA.

Note we do not consider the Bayes Factors for model comparisons here as they cannot be computed for the deep generative models.

\subsection{Nonlinear Factor Analysis} \label{sec:nonlinear-example}

Now consider a simulation setting where the true mapping from the factors to the observed data is nonlinear. In this situation, we expect that both the DGM and the skip-DGM will reconstruct the data well while PPCA will require a larger number of latent dimensions to model the nonlinear mapping. 

We set the number of samples to  $N=1000$, the number of observed features to $G=7$ and the latent dimension to $K=2$. The data is generated from
\begin{align}
	\mbx_i = (7z_{i1},\ 6z_{i1},\ 5z_{i1}^2, \ 4z_{i2},\ 3z_{i2},\ 2\sin(\pi/2 \cdot z_{i2}),\ 1z_{i1} \cdot z_{i2})^{\top} + \boldsymbol{\epsilon}_i, \label{eq:nonlinear}
\end{align}
where $\bm{z}_i \sim N(0, \bm{I})$, $\bm{\varepsilon}_i\sim N(0, \sigma^2\bm{I})$ with true $\sigma^2=1$.  That is, the first three columns of $\mbx$ are related to the first factor; the next three columns are related to the first factor; and the final column is an interaction term between the two factors.  Both the DGM and skip-DGM should be able to reconstruct the data using $K=2$ dimensions, whereas PPCA would need at least $K=5$ latent dimensions to capture the nonlinear terms. 

For the DGM and skip-DGM, we use a similar neural network architecture as the previous section, except with 50 neurons in each hidden layer. For PPCA, we consider both a model with the true number of latent dimensions, $K=2$, and a model with an overestimate of the latent dimension, $K = 5$. 

We first check each model. PPCA with $K=2$ fails the partial predictive check, as expected; two-dimensions are inadequate for a linear model to capture the data.  Each of PPCA $(K=5)$, the DGM and skip-DGM pass their partial predictive check (\Cref{fig:nonlinear-ppn}).  To assess which model (or set of models) to proceed with, we use the PPN study. Consider  \Cref{fig:nonlinear-ppn}:
\begin{itemize}
	\item (row 2, column 3): The PPN comparing PPCA-5 and the DGM (with the  PPCA-5 diagnostic) suggests that the DGM can fool PPCA-5.
	\item (row 2, column 3): The PPN comparing  PPCA-5  and the skip-DGM (with the  PPCA-5  diagnostic) suggests that the DGM can fool PPCA-5. 
	\item (row 3, column 2): The PPN comparing  PPCA-5  and the DGM (with the DGM diagnostic) suggests that PPCA-5 cannot fool the DGM. 
	\item (row 3, column 4): The PPN comparing the DGM and the skip-DGM (with the DGM diagnostic) suggests that the skip-DGM can fool the DGM.
	\item (row 4, column 2): The PPN comparing  PPCA-5  and the skip-DGM (with the skip-DGM diagnostic) suggests that the skip-DGM can fool PPCA-5.
	\item (row 4, column 3): The PPN comparing the DGM and the skip-DGM (with the skip-DGM diagnostic) suggests that the DGM can fool the skip-DGM.
\end{itemize}

The skip-DGM fools PPCA-5 , but  PPCA-5 does not fool the skip-DGM. This result suggests that the skip-DGM captures aspects of the data that PPCA-5 does not.  Based on the overlap of the posterior predictive distributions, both the DGM and skip-DGM fool each other, suggesting that they both capture the same aspects of the data.

\noindent
\begin{minipage}{\textwidth}

\begin{minipage}[c]{\textwidth}
\centering
\includegraphics[width =0.55\textwidth]{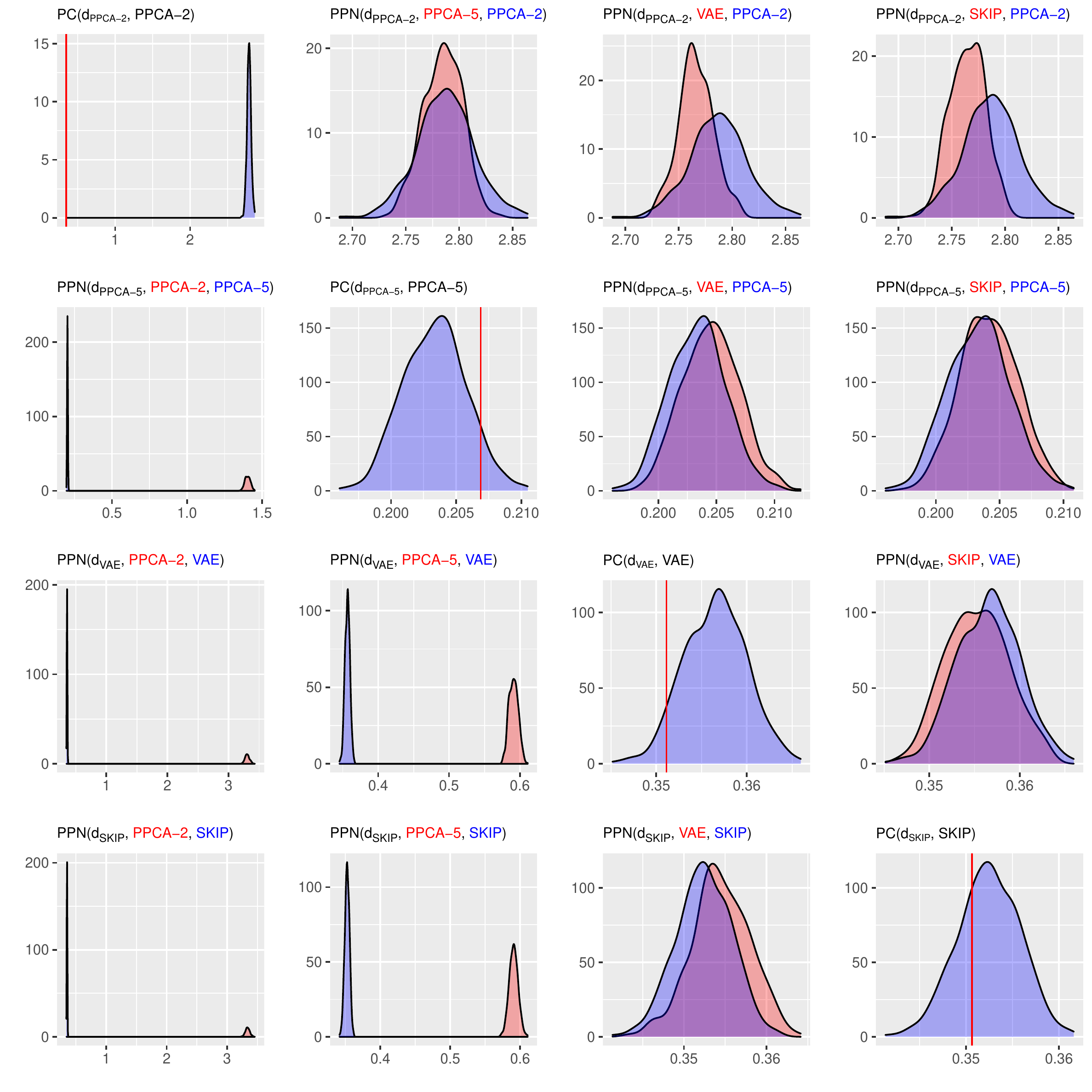}
\captionof{figure}{A PPN study of factor analysis models. This study suggests that a nonlinear model is consistent with the data (a linear model is not enough). On the diagonal are heldout predictive checks displaying $p_{\rmA}(d_A(\mbx_{\mathrm{rep}}^\rmA;\mbx_{\mathrm{val}})|\mbx_{\mathrm{in}})$ (blue histogram) and $d_A(\mbx_{\mathrm{out}};\mbx_{\mathrm{val}})$ (red line). PPCA-5, VAE and the skip-VAE pass their model checks. Beside the diagonal are PPNs, each one checking if a different model can fool the model under study. Specifically, the PPNs compare $p_{\rmB}(d_A(\mbx_{\mathrm{rep}}^\rmB;\mbx_{\mathrm{val}})|\mbx_{\mathrm{in}})$ (red histogram) and $p_{\rmA}(d_A(\mbx_{\mathrm{rep}}^\rmA;\mbx_{\mathrm{val}})|\mbx_{\mathrm{in}})$ (blue histogram).  The PPN study shows that PPCA-5 does not fool the skip-VAE and VAE, while the skip-VAE and VAE fool each other.}\label{fig:nonlinear-ppn}
\end{minipage}

\begin{minipage}[c]{\textwidth}
\centering
{\footnotesize
	\captionof{table}{Symmetrized Kullback-Leibler distance between distributions for the PPN study in \Cref{fig:nonlinear-ppn}. This study suggests that a nonlinear model is consistent with the data.} 
		\begin{tabular}{ l r  r r r }
		\toprule
& PPCA-2 & PPCA-5 & VAE & SKIP-VAE \\ 
\midrule
PPCA-2 &  & \cellcolor{gray!10}0.43 &\cellcolor{gray!30} 1.59 &\cellcolor{gray!30}1.72\\
PPCA-5 &\cellcolor{gray!70} 12.44 &  &\cellcolor{gray!10} 0.37 &\cellcolor{gray!10} 0.26\\
VAE& \cellcolor{gray!70}12.68 & \cellcolor{gray!70}14.31 &  &\cellcolor{gray!10} 0.26\\
SKIP-VAE &\cellcolor{gray!70} 12.21 &\cellcolor{gray!70}14.31 &\cellcolor{gray!10} 0.47 & \\
\bottomrule
		\end{tabular}}
\end{minipage}
\end{minipage}


\section{Discussion}
\label{sec:conclusion}

We developed and studied the posterior predictive null check (PPN), an
approach to Bayesian model criticism that complements the classical
predictive checks.  A PPN checks whether data from model $\rmB$'s
posterior predictive distribution can pass the predictive check of
model $\rmA$.  By studying a space of models with a collection of
PPNs, we can understand the relationships between them. Which models capture different aspects of the data?

With mixtures, we demonstrated how a PPN study can help select a model
by the principle of parsimony. With probabilistic factor models, we
demonstrated how it can help understand relationships between
different classes of models.  We re-analyzed data from the research
literature on Bayesian model criticism, and we studied the calibration
properties of the PPN.

{\edits Running a PPN study is more computationally expensive than computing predictive checks. This expense is because for $M$ models, a PPN study considers order $M^2$ model combinations. This computational expense may be mitigated when the models are ordered by complexity. In this case, a PPN study can proceed by comparing only consecutive models (i.e. $\mathcal{M}_k$ vs. $\mathcal{M}_{k+1}$), reducing the number of model combinations to $M$.
}

In the modern practice of applied Bayesian statistics, researchers
iteratively design and explore many models, a process that was
recently dubbed ``the Bayesian workflow'' \citep{gelman2020bayesian}.
By helping the researcher understand the relationships between
different models, and particularly so in the context of Bayesian model
criticism, a PPN study can help guide the researcher through this
process.

{\edits
All of the PPN studies here, both with real data and simulated data, involve a situation where more than one model passes its predictive check.  One interesting direction of future work is to consider a PPN study where no model passes its check, but where we might still be interested in understanding the differences between the models' predictive distributions.
}



\bibliographystyle{ba}
\bibliography{bib/references}

\appendix

\section{Gaussian mixture model} \label{appendix:gmm}

Here, we provide details regarding the Gaussian mixture model in \Cref{sec:intro} and \Cref{sec:ppn-ppn} of the main paper.  The data $\mbx_i \in \R^D$, $i = 1,\dots, N$ is generated from the following model with $D=2$, $K^*=3$ and $N=500$.
\begin{align}
\bm{\gamma}_i &\sim \text{Multinomial}(\bm{p}, 1), \quad \bm{p} = \{1/K^*\}_{k=1}^{K^*}\\
\mbx_i |\bm{\gamma}_i, \bm{\mu}, \bm{\Sigma} &\sim \sum_{k=1}^{K^*} \gamma_{ik}N(\bm{\mu}_k, \bm{\Sigma}_k), \quad \text{where}\quad \bm{\Sigma}_k = \text{diag}\{\sigma_{k1}^2, \dots, \sigma_{kd}^2\},
\end{align}
with 
\begin{align}
\{\bm{\mu}_{dk}\}_{d, k=1}^{D, K^*} =
\begin{pmatrix}
-5 & 0 & 10 \\
5 & 0 & 5
\end{pmatrix} 
\qquad \text{and} \qquad
\{\sigma^2_{dk}\}_{d, k = 1}^{D, K^*} = 
\begin{pmatrix}
1 & 2 & 2 \\
1 & 1 & 4
\end{pmatrix}.
\end{align}

We fit the model:
\begin{align}
\bm{\gamma}_i &\sim \text{Multinomial}(\bm{p}, 1), \quad \bm{p} = \{1/K\}_{k=1}^K \\
\sigma_{kd}^2 &\sim \text{Inverse-Gamma}(1, 1) \\
\bm{\mu}_k &\sim N(0, 5^2) \\
\mbx_i |\bm{\gamma}_i, \bm{\mu}, \bm{\Sigma} &\sim \sum_{k=1}^K \gamma_{ik}N(\bm{\mu}_k, \bm{\Sigma}_k).
\end{align}

Denote $\bm{\theta} = \{\bm{\mu}_k, \bm{\Sigma}_k\}_{k=1}^K$.  We set the true number of components to be $K^*=3$. We run a PPN check to compare models with $K \in \{1,2,3,4\}$ components.  We fit each of these models using Gibbs sampling. 

We conduct the partial predictive and PPN checks with the heldout diagnostic \eqref{eq:heldout-diagnostic}. The underlying realized diagnostic function is the log-likelihood, given by 
\begin{align}
d_K(\mbx, \bm{\theta}) =  \sum_{k=1}^K \left\{\sum_{i=1}^n \left\{ -\frac{1}{2}\gamma(\mbx_i, \bm{\theta}) (\mbx_i - \bm{\mu}_k)^T\bm{\Sigma}_k(\mbx_i - \bm{\mu}_k)\right\} - \frac{1}{2}\sum_{i=1}^n\gamma(\mbx_i, \bm{\theta}) \log|\bm{\Sigma}_k|\right\}
\end{align}
where $\gamma(\mbx_i, \bm{\theta})$ is a draw from $p(\gamma|\mbx_i, \bm{\theta})$.  For both the partial predictive and PPN checks, we use $R=200$ draws from the posterior predictive distributions.

\section{Comparison to Bayes Factors}\label{appendix:bayes_factors}

To compare two models, $\mathcal{M}_{\mathrm{A}}$ and $\mathcal{M}_{\mathrm{B}}$, the Bayes factor is
\begin{align}
B_{AB} = \frac{p(\mbX|\mathcal{M}_{\mathrm{A}}) p(\mathcal{M}_{\mathrm{A}})}{p(\mbX|\mathcal{M}_{\mathrm{B}}) p(\mathcal{M}_{\mathrm{B}})}.
\end{align}
In the mixture model example, we approximate the marginal likelihood, $p(\mbX|\mathcal{M})$, with the harmonic mean of the likelihood values \citep{newton1994approximate}:
\begin{align}
\mathrm{p}(\mbX|\mathcal{M}) = \left\{\frac{1}{R}\sum_{r=1}^R P(\mbX|\theta^{(r)}, \mathcal{M})^{-1}\right\}^{-1}, \label{eq:harmonic_mean}
\end{align}
where $\theta^{(r)}\sim p(\theta|\mbX, \mathcal{M})$ are draws from the posterior under model $\mathcal{M}$.

For the Gaussian mixture model example, the Bayes factors suggest $K=3$, similarly to the PPN (\Cref{table:bf_gaussian}).

For the multinomial mixture model example, the Bayes factors provide inconclusive evidence (\Cref{table:multinomial_bf}).

\begin{table}[ht]
\centering
\caption{Bayes Factors for Gaussian mixture model example.}\label{table:bf_gaussian}
\begin{tabular}{rrrrr}
  \toprule
$\mathcal{M}_{\mathrm{A}} \backslash \mathcal{M}_{\mathrm{B}}$ & $K=1$ & $K=2$& $K=3$ & $K=4$ \\ 
  \midrule
$K=1$ &  & &  &  \\ 
  $K=2$ & 3.19 & &  &  \\ 
  $K=3$ & 11.52 & 3.62 & &  \\ 
  $K=4$ & 11.97 & 3.76 & 1.04 & \\ 
   \bottomrule
\end{tabular}
\end{table}

\begin{table}[ht]
\centering
\caption{Bayes Factors for the multinomial mixture model.} \label{table:multinomial_bf}
\begin{tabular}{rrrrr}
  \toprule
$\mathcal{M}_{\mathrm{A}} \backslash \mathcal{M}_{\mathrm{B}}$ & $K=1$ & $K=2$& $K=3$ & $K=4$ \\ 
  \midrule
$K=1$ & & 1.37 & 1.84 & 2.23 \\ 
 $K=2$ & 0.73 & & 1.35 & 1.63 \\ 
  $K=3$ & 0.54 & 0.74 & & 1.21 \\ 
  $K=4$& 0.45 & 0.61 & 0.83 &\\ 
   \bottomrule
\end{tabular}
\end{table}

\newpage
\section{Proof of Proposition 1}\label{appendix:regression}

The PPN compares the distributions: $\rmp_\rmA (d_\rmB(\bm{y}_{\mathrm{rep}}^\rmA;\bm{y}_{\mathrm{val}})\g\bm{y}_\mathrm{in}, \mbX_{\mathrm{in}})$ and $\rmp_\rmB (d_\rmB(\bm{y}_{\mathrm{rep}}^\rmB;\bm{y}_{\mathrm{val}})\g\bm{y}_\mathrm{in}, \mbX_{\mathrm{in}})$.  The posterior predictive distribution of model $\rmA$ is:
\begin{align}
y_{\mathrm{rep}, i}^\rmA | \bm{y}_{\mathrm{in}}\sim N(\overline{\bm{y}}_{\mathrm{in}}, 2). \label{eq:post_pred_regression_A}
\end{align}
The posterior predictive distribution of model $\rmB$ is:
\begin{align}
y_{\mathrm{rep},i}^\rmB | \bm{y}_{\mathrm{in}} \sim N(\widehat{\theta}_{\mathrm{in}}^\rmB + \mbx_{\mathrm{in}, i}^{\top}\widehat{\beta}_{\mathrm{in}}^\rmB, 2 + \mbx_{\mathrm{in}, i}^{\top}(\mbX_{\mathrm{in}}^{\top}\mbX_{\mathrm{in}})^{-1}\mbx_{\mathrm{in}, i}), \label{eq:post_pred_regression_B}
\end{align}
where $\widehat{\beta}_{\mathrm{in}}^\rmB = (\mbX_{\mathrm{in}}^{\top}\mbX_{\mathrm{in}})^{-1}\mbX_{\mathrm{in}}^{\top}\bm{y}_{\mathrm{in}}$ and $\widehat{\theta}_{\mathrm{in}}^\rmB = \overline{\bm{y}}_{\mathrm{in}} -\overline{\mbX}_{\mathrm{in}}\widehat{\beta}_{\mathrm{in}}$.

We first consider the distribution of data from model $\rmA$ under the model $\rmB$ diagnostic:
\begin{align}
d_\rmB(\bm{y}_{\rep}^\rmA;\bm{y}_{\mathrm{val}}) &= \sum_{i=1}^n [y_{\rep,i}^\rmA -\widehat{\theta}_{\mathrm{val}}^\rmB - \mbx_{\mathrm{in}, i}^\top\widehat{\bm{\beta}}_{\mathrm{val}}^\rmB ]^2 \\
&=\sum_{i=1}^n [y_{\rep, i}^\rmA - \overline{\bm{y}}_{\mathrm{in}} - (\widehat{\theta}_{\mathrm{val}}^\rmB -\overline{\bm{y}}_{\mathrm{in}})- \mbx_{\mathrm{in}, i}^\top(\widehat{\bm{\beta}}_{\mathrm{val}}^\rmB - \bm{\beta}_0)]^2 
\end{align}
where $\bm{\beta}_0=\bm{0}_p$ is the true parameter value. As $y_{\rep, i}^\rmA -\overline{\bm{y}}_{\mathrm{in}} \sim N(0, 2)$, we have:
\begin{align}
d_\rmB(\bm{y}_{\rep}^\rmA;\bm{y}_{\mathrm{val}})&= \sum_{i=1}^n [\sqrt{2}\cdot Z_i - (\widehat{\theta}_{\mathrm{val}}^\rmB -\overline{\bm{y}}_{\mathrm{in}})- \mbx_{\mathrm{in}, i}^\top(\widehat{\bm{\beta}}_{\mathrm{val}}^\rmB - \bm{\beta}_0)]^2, 
\end{align}
where $Z_i \stackrel{ind}{\sim}N(0, 1)$.  As $n\to\infty$, we have:
\begin{align}
(\widehat{\theta}_{\mathrm{val}}^\rmB -\theta_0 -(\overline{\bm{y}}_{\mathrm{in}}-\theta_0))\sim N(0, 2/n).
\end{align}
We have
\begin{align}
\mbx_{\mathrm{in}, i}^\top(\widehat{\bm{\beta}}_{B, \mathrm{val}} - \bm{\beta}_0)\sim N\left(0, \mbx_{\mathrm{in}, i}^\top[\mbX_{\mathrm{val}}^\top\mbX_{\mathrm{val}}]^{-1}\mbx_{\mathrm{in}, i} \right).
\end{align}
By \Cref{eq:covariate_condition}, we have $\mbx_{\mathrm{in}, i}^\top[\mbX_{\mathrm{val}}^\top\mbX_{\mathrm{val}}]^{-1}\mbx_{\mathrm{in}, i} \to p/n$ as $n\to\infty$. 

Then, as $n\to\infty$ and $p/n\to 0$, $d_B(\bm{y}_A)$ has an approximately $2\chi_n^2$-distribution (to first order).

We now consider the distribution of data from $\mathcal{M}_B$ under the $\mathcal{M}_B$ diagnostic: 
\begin{align}
d_\rmB(\bm{y}_{\rep}^\rmB;\bm{y}_{\mathrm{val}}) &= \sum_{i=1}^n [y_{\rep,i}^{\rmB} -\widehat{\theta}_{\mathrm{val}}^\rmB - \mbx_{\mathrm{in}, i}^\top\widehat{\bm{\beta}}_{\mathrm{val}}^\rmB ]^2  \\
&= \sum_{i=1}^n [y_{\rep,i}^{\rmB} -\widehat{\theta}_{\mathrm{in}}^\rmB - \mbx_{\mathrm{in},i}^\top\widehat{\bm{\beta}}_{\mathrm{in}}^\rmB + (\widehat{\theta}_{\mathrm{in}}^\rmB -\widehat{\theta}_{\mathrm{val}}^\rmB) + \mbx_{\mathrm{in}, i}^\top(\widehat{\bm{\beta}}_{\mathrm{in}}^\rmB- \widehat{\bm{\beta}}_{\mathrm{val}}^\rmB) ]^2 \\
&= \sum_{i=1}^n \left[\sqrt{2 + \mbx_{\mathrm{in}, i}^{\top}(\mbX_{\mathrm{in}}^{\top}\mbX_{\mathrm{in}})^{-1}\mbx_{\mathrm{in}, i}}\cdot\widetilde{Z}_i + (\widehat{\theta}_{\mathrm{in}}^\rmB -\widehat{\theta}_{\mathrm{val}}^\rmB) + \mbx_{\mathrm{in}, i}^\top(\widehat{\bm{\beta}}_{\mathrm{in}}^\rmB- \widehat{\bm{\beta}}_{\mathrm{val}}^\rmB) \right]^2 
\end{align}
where $\widetilde{Z}_i\sim N(0, 1)$.
As $n\to\infty$, we have:
\begin{align}
(\widehat{\theta}_{\mathrm{in}}^\rmB -\widehat{\theta}_{\mathrm{val}}^\rmB)\sim N(0, 2/n).
\end{align}
Further,
\begin{align}
\mbx_{\mathrm{in}, i}^\top(\widehat{\bm{\beta}}_{\mathrm{in}}^\rmB- \widehat{\bm{\beta}}_{\mathrm{val}}^\rmB) \sim N(0, \mbx_{\mathrm{in}, i}^\top[\mbX_{\mathrm{val}}^\top\mbX_{\mathrm{val}}]^{-1}\mbx_{\mathrm{in}, i} + \mbx_{\mathrm{in}, i}^\top[\mbX_{\mathrm{in}}^\top\mbX_{\mathrm{in}}]^{-1}\mbx_{\mathrm{in}, i}).
\end{align}
As $n\to\infty$, we have $\mbx_{\mathrm{in}, i}^\top[\mbX_{\mathrm{val}}^\top\mbX_{\mathrm{val}}]^{-1}\mbx_{\mathrm{in}, i} + \mbx_{\mathrm{in}, i}^\top[\mbX_{\mathrm{in}}^\top\mbX_{\mathrm{in}}]^{-1}\mbx_{\mathrm{in}, i}\to 2p/n$.

Then, as $n\to\infty$ and $p/n\to 0$, we have $d_\rmB(\bm{y}_{\rep}^\rmB;\bm{y}_{\mathrm{val}})$ has an approximately $2\chi_n^2$-distribution (to first order).

Thus, $d_\rmB(\bm{y}_{\rep}^\rmB;\bm{y}_{\mathrm{val}})$ is asymptotically equal in distribution to $d_\rmB(\bm{y}_{\rep}^\rmA;\bm{y}_{\mathrm{val}})$ (to first order).

 \begin{acknowledgement}
This research was supported by ONR N00014-17-1-2131, ONR N00014-15-1-2209, DARPA SD2 FA8750-18-C-0130, the Simons Foundation, NSF NeuroNex, the Sloan Foundation, the McKnight Endowment, and the Gatsby Charitable Trust.  

We thank Scott Linderman for helpful discussions about this work. 
\end{acknowledgement}

\end{document}